\documentclass{aa}  

\usepackage{graphicx}
\usepackage{txfonts}
\usepackage{multirow}
\usepackage{color}
\usepackage{amssymb}
\usepackage{lscape}
\usepackage{longtable}
\usepackage{epsfig}
\usepackage{url}
\usepackage{natbib,twoopt}
\newcommand{\hii}{H\,{\sc{ii}}}
\newcommand{\mm}{\,\mu{\rm m}}

\newcommand{\spitzer}{{\it Spitzer}}
\newcommand{\herschel}{{\it Herschel}}
\newcommand{\mia}{RCW~120}

\newcommand{\co}{$^{12}$CO}
\newcommand{\tco}{$^{13}$CO}
\newcommand{\pdr}{CLay}
\newcommand{\av}{A_{{\rm v}}}
\newcommand{\nh}{n_{{\rm H}}}
\newcommand{\Nh}{N_{{\rm H}}}
\newcommand{\dpdr}{\Delta pdr}
\newcommand{\cir}{CIr}
\newcommand{\cicotr}{CICOtr}
\newcommand{\rco}{COr}
\newcommand{\gra}{GrA}
\newcommand{\grb}{GrB}
\newcommand{\grc}{GrC}
\newcommand\T{\rule{0pt}{2.6ex}}
\newcommand\B{\rule[-1.5ex]{0pt}{0pt}}

\begin{document} 

	\title{$\it{Herschel}$\thanks{\herschel\ is an ESA space observatory with science instruments
provided by European-led Principal Investigator consortia and with important
participation from NASA.} SPIRE-FTS observations of RCW~120}

	\author{Rod\'on, J.A.\inst{1,2}
		\and
		Zavagno, A.\inst{2}
		\and
		Baluteau, J.-P.\inst{2}
		\and  
		Habart, E.\inst{3}
		\and
		K\"ohler, M.\inst{3}
		\and
		Le Bourlot, J.\inst{4}
		\and
		Le Petit, F.\inst{4}
		\and
		Abergel, A.\inst{3}
}%

	\institute{European Southern Observatory, Alonso de C\'ordova 3107, Vitacura, Santiago 19, Chile.\\
	\email{jrodon@eso.org}
	\and
	Aix Marseille universit\'e, CNRS, LAM (Laboratoire d'Astrophysique de Marseille) UMR 7326, 13388 Marseille, France.
	\and 
	Institut d'Astrophysique Spatiale, CNRS/Universit\'e Paris-Sud\,11, 91405 Orsay, France 
	\and
	LUTH, Observatoire de Paris et Universit\'e Paris 7, 5 Place Jules Janssen, F-92190 Meudon, France \\
	} %

   \date{Received , accepted }

 
\abstract
{The expansion of Galactic H\,{\sc{ii}} regions can trigger the formation of a new generation of stars. However, little is know about the physical conditions that prevail in these regions. }
{We study the physical conditions that prevail in specific zones towards expanding \hii\ regions that trace representative media such as the photodissociation region, the ionized region, and condensations with and without ongoing star formation.}
{We use the SPIRE Fourier Transform Spectrometer (FTS) on board \herschel\ to observe the \hii\ region \mia. Continuum and lines are observed in the $190-670\mm$ range. Line intensities and line ratios are obtained and used as physical diagnostics of the gas. We used the Meudon PDR code and the RADEX code to derive the gas density and the radiation field at nine distinct positions including the PDR surface and regions with and without star-formation activity.}
{For the different regions we detect the atomic lines [NII] at $205\mm$ and [CI] at $370$ and $609\mm$, the \co\ ladder between the $J=4$ and $J=13$ levels and the \tco\ ladder between the $J=5$ and $J=14$ levels, as well as CH$ ^{+} $ in absorption.
We find gas temperatures in the range $45-250\,$K for densities of $10^4-10^6\,{\rm cm}^{-3}$, and a high column density on the order of $\Nh\sim10^{22}\,{\rm cm}^{-2}$ that is in agreement with dust analysis.
The ubiquitousness of the atomic and CH$ ^{+} $ emission suggests the presence of a low-density PDR throughout \mia. High-excitation lines of CO indicate the presence of irradiated dense structures or small dense clumps containing young stellar objects, while we also find a less dense medium ($\Nh\sim10^{20}\,{\rm cm}^{-2}$) with high temperatures ($80-200\,$K).}
{}


   \maketitle
%

\section{Introduction}

Star formation occurs on the borders of Galactic \hii\ regions. Different physical processes may be at work to trigger this star formation \citep{elmegreen1977,deharveng2010}.
This phenomenon has been studied in detail over the past ten years \citep{deharveng2003,zavagno2007,pomares2009,zavagno2010,paron2011,davies2012}. All these studies concentrate on specific Galactic \hii\ regions where young stellar objects (YSOs) are observed on their borders. This phenomenon is also observed in nearby galaxies (see, e.g., the case of N11 in the Large Magellanic Cloud).   The \spitzer\ satellite and the GLIMPSE and MIPSGAL surveys of the Galactic Plane have revealed that we are living in a bubbling galactic disk where \hii\ regions have a clear impact on their environment.
A study using \spitzer\ GLIMPSE and MIPSGAL data combined with ATLASGAL data on 102 bubbles have shown that the star formation triggered by \hii\ regions is an important phenomenon in our Galaxy. Up to 25\% of the bubbles show triggered massive-star formation on their border \citep{deharveng2010}.

The \herschel\ satellite offers a unique opportunity to study the star formation triggered by Galactic \hii\ regions. Thanks to its sensitivity and its large wavelength coverage in the far infrared, \herschel\ is perfectly suited to the study of the earliest phases of star formation.

We are engaged in several guaranteed (HOBYS, \citealt{motte2010}, ``Evolution of Interstellar Dust'', \citealt{abergel2010}) and open time (Hi-GAL, \citealt{molinari2010}) key programs on \herschel\ that aim to characterize this way of forming stars. We used PACS and SPIRE photometers to characterize the emission in the $60-500\mm$ range towards Galactic \hii\ regions. This allows us to detect and characterize the properties of young stellar objects (YSOs) observed towards these regions. We also used the PACS and SPIRE spectrometers to derive the physical conditions towards these regions.

Here we present results of \herschel\ SPIRE-FTS spectroscopy obtained towards the Galactic \hii\ region \mia. This region is among the closest \hii\ regions with a distance of $ \sim1.3\,$kpc \citep{russeil2003}, which combined to its angular diameter of $\sim7'.5$ \citet{anderson2015} results in a physical size of $ \sim3 $pc.
\mia\ is ionized by the single star $ CD-38^{\circ}11636 $, with a spectral type O6-8V/III according to the latest measurements by \citet{martins2010}.

\begin{figure*}[th!]
	\centering
	\includegraphics[width=\textwidth]{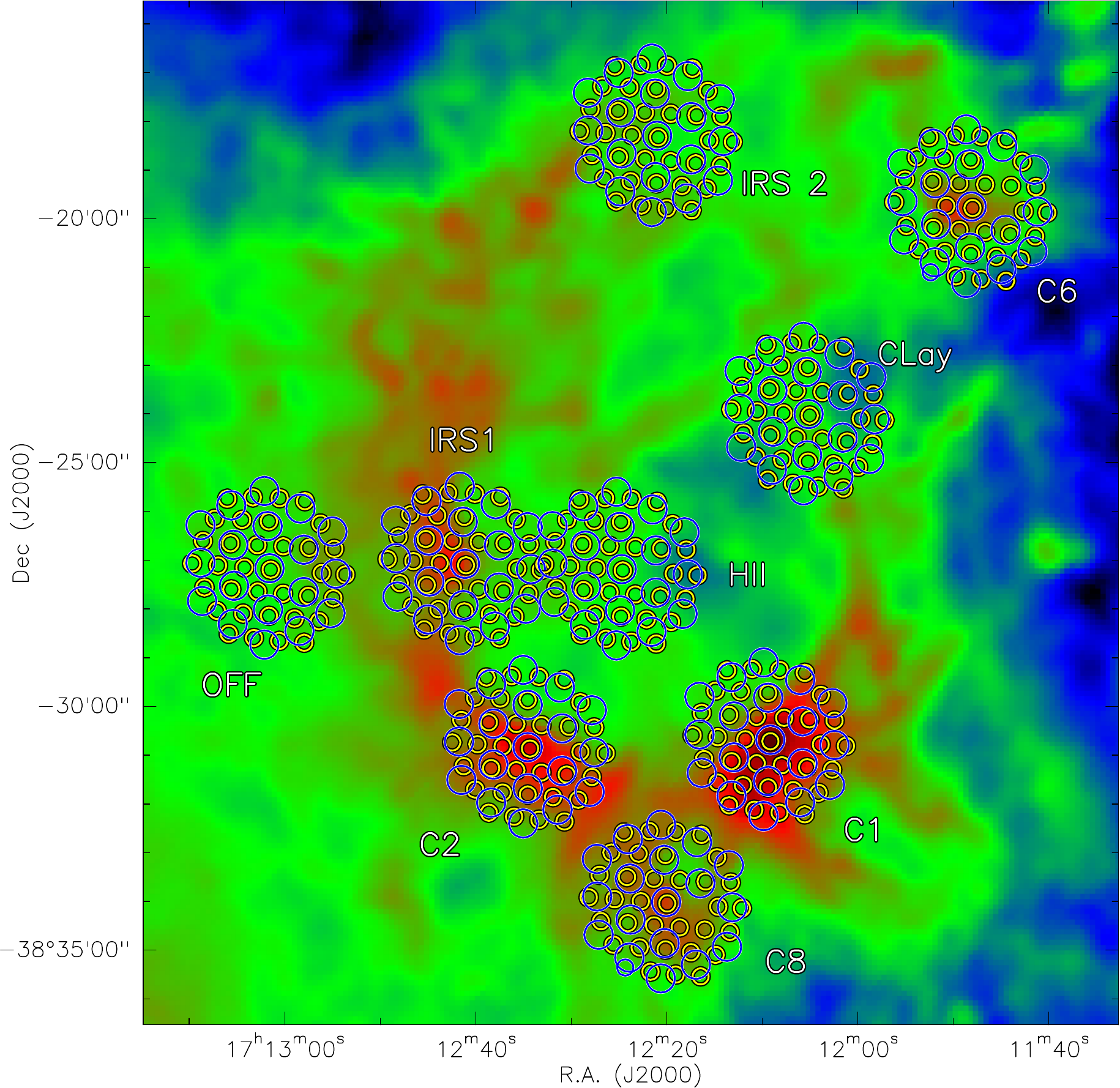}
	\caption{\herschel\ SPIRE-FTS pointing towards RCW~120: \herschel\ SPIRE $350\mm$ image (background) of RCW~120 on which the SLW (blue) and SSW detectors (yellow) are superimposed. The nine pointings observed are labeled according to their location (see text). }
	\label{fig-pointings}
\end{figure*}

Labeled ``the perfect bubble'' throughout the literature, recent single-dish observations of the lowest CO transitions \citep{anderson2015,torii2015} fail to detect an expanding shell of fore- and background material, which would indicate a 3-D structure. However, \citet{anderson2015} finds discrete ``holes'' in the PDR, through which the ionizing radiation is escaping, and \citet{torii2015} propose an explanation of the observed morphology of \mia\ through the collision of two clouds, which formed the ionizing star.

Observations in mm and sub-mm wavelengths \citep{zavagno2007,deharveng2009} show a fragmented neutral layer along its PDR, which contains five of the eight mm-condensations found by \citet{zavagno2007}. According to \citet{deharveng2009} the mass of the layer is $ \sim2000\,$M$_{\sun}$.
The most massive condensation has been resolved into a chain of several Class~I objects \citep{deharveng2009}, likely an example of Jeans instability, while \citet{zavagno2010a} found a massive $ 8-10\, $M$ _{\sun} $ YSO towards this condensation, which they suggest is 
the first detection of a massive Class~0 object formed by the collect and collapse process on the border of an \hii\ region.

\begin{table*}[]
	\renewcommand{\arraystretch}{1.2}  
    	\centering
    	\caption{SPIRE-FTS observations on Operational Day 288 (February 26 2010).}
    	\label{table-obs}
    	\renewcommand{\footnoterule}{}
		\begin{tabular}{l|cc|cccc}
			\hline\hline			
			Pointing & RA (J2000) & Dec (J2000) & Obs. ID & Start time (UT) & Int. time (s) & Repetitions \\
			\hline             
			C6 & 17:11:47.89 & $-$38:19:48.2 & 1342191225 & 15:21:40.00 & 1353 & 8 \\
			\pdr\ & 17:12:04.91 & $-$38:24:03.5 & 1342191226 & 15:44:38.00 & 1353 & 8 \\
			C1 & 17:12:08.93 & $-$38:30:43.1 & 1342191227 & 16:07:38.00 & 543 & 2 \\
			C8 & 17:12:19.81 & $-$38:34:04.0 & 1342191228 & 16:17:04.00 & 543 & 2 \\
			IRS2 & 17:12:20.77 & $-$38:18:21.1 & 1342191229 & 16:26:43.00 & 1353 & 8 \\
			HII & 17:12:24.47 & $-$38:27:14.3 & 1342191230 & 16:49:45.00 & 543 & 2 \\
			C2 & 17:12:34.15 & $-$38:30:51.9 & 1342191231 & 16:59:11.00 & 543 & 2 \\
			IRS1 & 17:12:40.89 & $-$38:27:07.9 & 1342191232 & 17:08:37.00 & 543 & 2 \\
			OFF & 17:13:01.35 & $-$38:27:13.3 & 1342191233 & 17:18:03.00 & 1353 & 8 \\
			\hline\hline
		\end{tabular}
	\tablefoot{Coordinates are for the central pixels of the SLW and SSW arrays (pixels C3 and D4, repectively).}
\end{table*}

\section{Observations and data reduction}
\label{obs}

\begin{figure}[h]
	\includegraphics[width=0.242\textwidth]{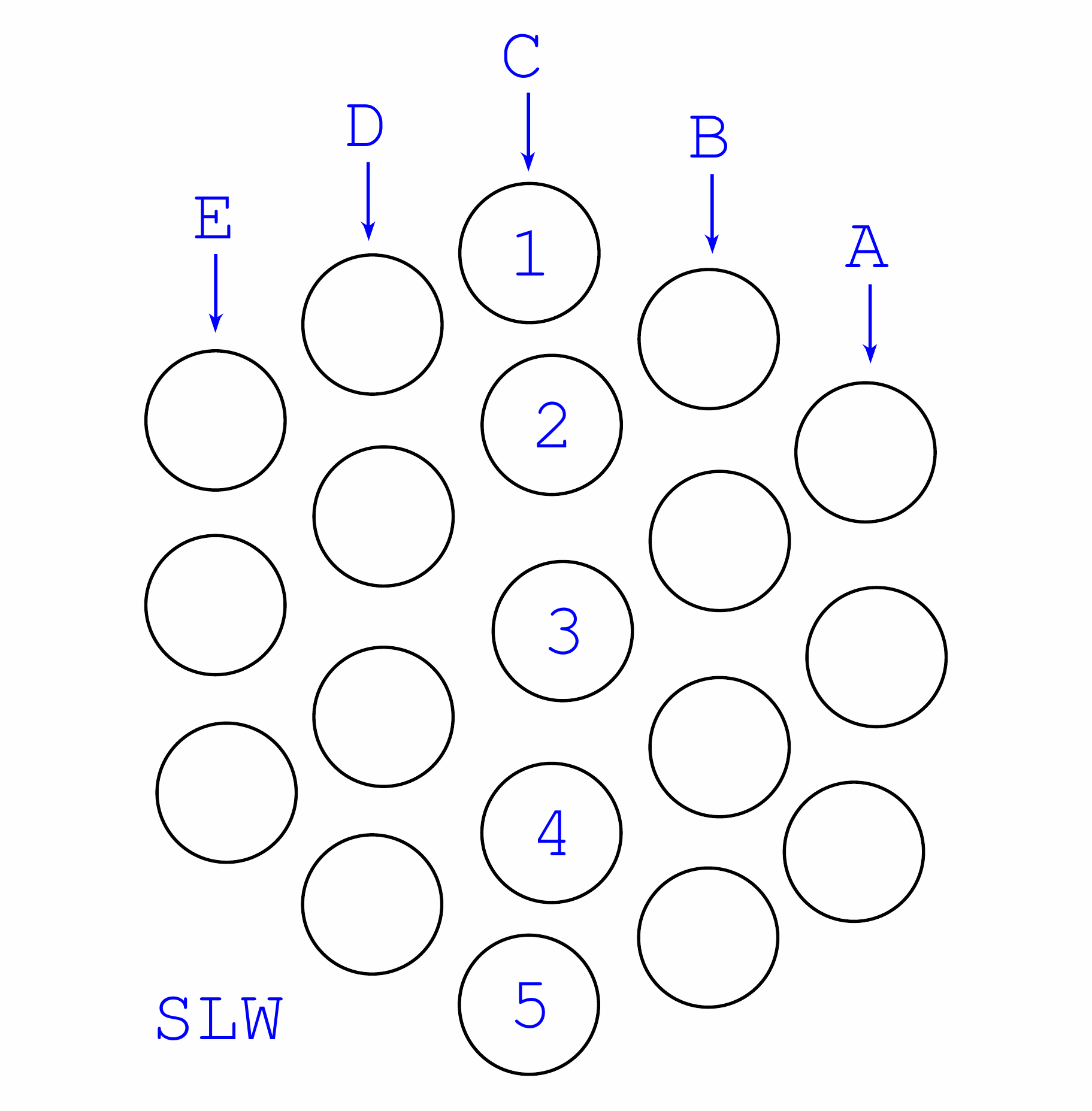}
	\includegraphics[width=0.242\textwidth]{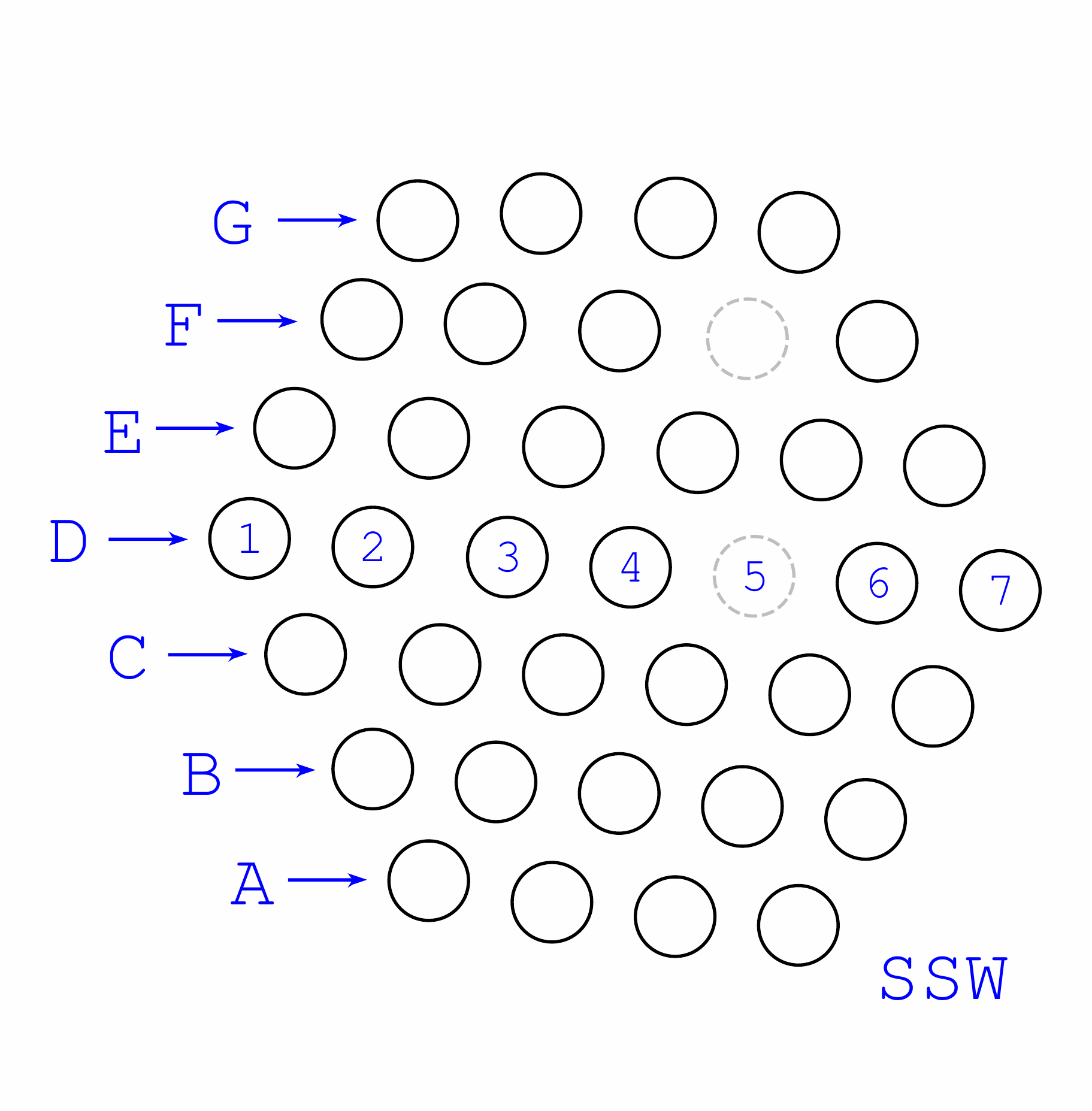}
	\caption{Disposition of the detectors (pixels) on the SSW (right) and SLW (left) arrays of the SPIRE-FTS instrument for our observations as seen projected in the sky. The two gray pixels in the SSW array mark dead pixels.}
	\label{fig-detectors}
\end{figure}

\mia\ was observed in spectroscopy with the SPIRE-Fourier Transform Spectrometer (SPIRE-FTS, \citealt{griffin2010}) on 26 February 2010 (OD 288), as part of the \textit{Evolution of Interstellar Dust} key program \citep{abergel2010}\footnote{The reduced data cubes are available on the HESIOD portal (\url{http://idoc-herschel.ias.u-psud.fr/sitools/client-user/})}.
The aim of the program is to study the physical conditions that exist in regions of triggered star formation.
For this purpose we selected 8 representative positions towards \mia: towards the ionized gas (pointing HII),
towards the photo-dissociation region (PDR) without any condensation or star formation (pointing \pdr), and towards condensations with ongoing star formation (see Sec. \ref{sec-pointings}). In \mia, nine condensations suspected of harboring ongoing star formation were observed in the mm continuum by \citet{zavagno2007}. For the FTS observations, we selected six such condensations, labeled C1, C2, C6, C8, IRS1 and IRS2. A position ``off'' the \hii\ region was also observed (pointing OFF). All nine pointings are shown in Fig.~\ref{fig-pointings}. The short wavelength (SSW, $194-320\mm$) and long wavelength (SLW, $313-671\mm$) arrays of the SPIRE-FTS are overlaid on the $350\mm$ SPIRE image of \mia, in yellow and blue, respectively. The arrays are composed of an hexagonal grid of $19$ pixels for SLW, and $37$ pixels for SSW. The pixel distribution and labels are shown in Fig.~\ref{fig-detectors}. There are two ``dead'' pixels in SSW, pixels F4 and D5, and are thus not shown in Fig.~\ref{fig-pointings} (missing yellow circles).

All the observations were performed in the high-resolution mode of the SPIRE-FTS. In general two scan repetitions were done, except for positions expected to be fainter, where eight scan repetitions were taken.
An observation summary is given in Table~\ref{table-obs}.

The data reduction was performed using the \herschel\ Interactive Processing Environment (HIPE) version 9.0 build 3048 with the calibration tree 9\_1. An iterative spectral line fitting routine has been used to extract the line parameters (line center, intensity and associated errors) from the unapodized FTS spectra. The routine used the unapodized instrumental line shape (the classical \textit{sinc} function) as the lines were checked to be unresolved by the FTS. The two scan directions were treated separately. As a conservative approach, the retained final error is the maximum value between the fitting procedure errors and the uncertainties obtained from the two direction results.

\subsection{Pointings}
\label{sec-pointings}
The nine pointings were selected on the basis of what was known from the infrared and millimeter study of \mia\ by \citet{zavagno2007}.
Their $1.3\,$mm continuum emission map revealed the presence of nine condensations, five of which are located on the PDR surrounding the ionized region of \mia. 
The idea of this SPIRE-FTS study was to study the physical conditions that prevail in the different regions located around \mia, regions that sample different 
evolution stage of star formation or different typical media (ionized region, PDR).

One FTS pointing was obtained on each of the main condensations (C1, C2, C6, C8 and IRS1). The stellar content of these condensations is described in detail in \citet{zavagno2007} and in \citet{deharveng2009}. The properties of the point sources observed in this region with $Herschel$ PACS and SPIRE are described in \citet{zavagno2010}.
An off position (OFF) was obtained to serve as a reference.
A position was obtained towards the ionized gas at the center of the ionized region (HII). A pointing was obtained towards the PDR, on the northeastern side of the region (CLay). 
All these pointings are described in more detail below.

C1: the most dense and massive condensation observed on the borders of RCW120. 
This region contains a massive Class~0 source revealed with $Herschel$ \citep{zavagno2010} and a chain of young, lower mass stars \citep{deharveng2009}. 

C2: this condensation contains a bright Class~I source and is, as C1, also located on the southern (and densest) border of the PDR.

C6: this condensation is located in the northeastern part of the region but not in direct contact with the main ionizing front.
However, a deep H$\alpha$ image shows that the ionized gas leaking from the \hii\ region hits this zone and is in direct contact with it (see Fig.~3 in \citealt{deharveng2009}), possibly acting on the star formation there. This condensation is dense and contains many bright Class~I YSOs.

C8: this condensation is located in the southern part of the ionized region and has probably been shaped by the leaking UV field (see Fig.~16 in \citealt{deharveng2009}). This region contains YSOs seen by $Herschel$ and is a site of active star formation.

IRS1: this region is located in the west part of the PDR and is labeled condensation 4 in \citet{zavagno2007}. It was renamed IRS1 for the FTS pointing because this region hosts bright IR YSOs observed with \textit{Spitzer}, but the peak of the $1.3\,$mm map was devoid of sources (see Fig. 8 in \citealt{zavagno2007}). 
This region is of interest because it samples very different physical conditions on a small spatial scale.    

IRS2: this region is located on the northern part of the PDR. The central pixel position was chosen to match with the position of the source IRAS$\,17089-3814$.

CLay: with the idea of sampling all the different physical conditions that exist towards \mia, we selected this zone known to be free of condensations detected in the mm map and star formation. Therefore this region should be representative of the PDR. At the time of the selection, the only process envisioned 
to form the layer was the collect and collapse \citep{elmegreen1977}, therefore the name Collected Layer (CLay).

HII: this pointing is obtained towards the ionized region. No star formation and no $1.3\,$mm continuum emission is observed towards this region.

\begin{figure}
	\includegraphics[width=0.47\textwidth]{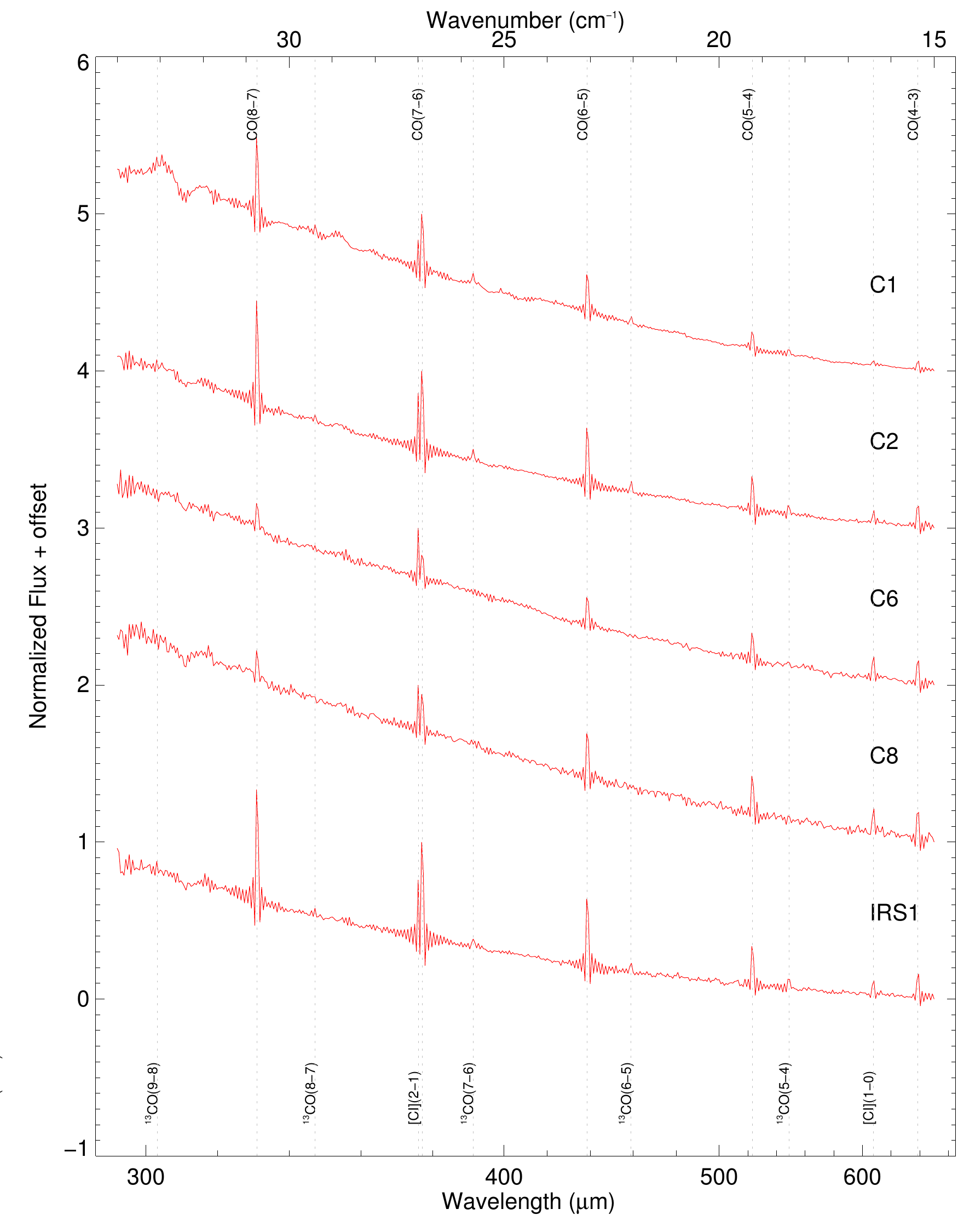}
	\caption{SPIRE-FTS SLW spectra towards the richest pointings, at the position of the central pixels (see Sec.~\ref{sec-res}). The spectra are normalized at the brightest line on each spectrum. The lines detected are marked and labeled.}
	\label{fig-spectra-lsw-h}
\end{figure}

\begin{figure}
	\includegraphics[width=0.47\textwidth]{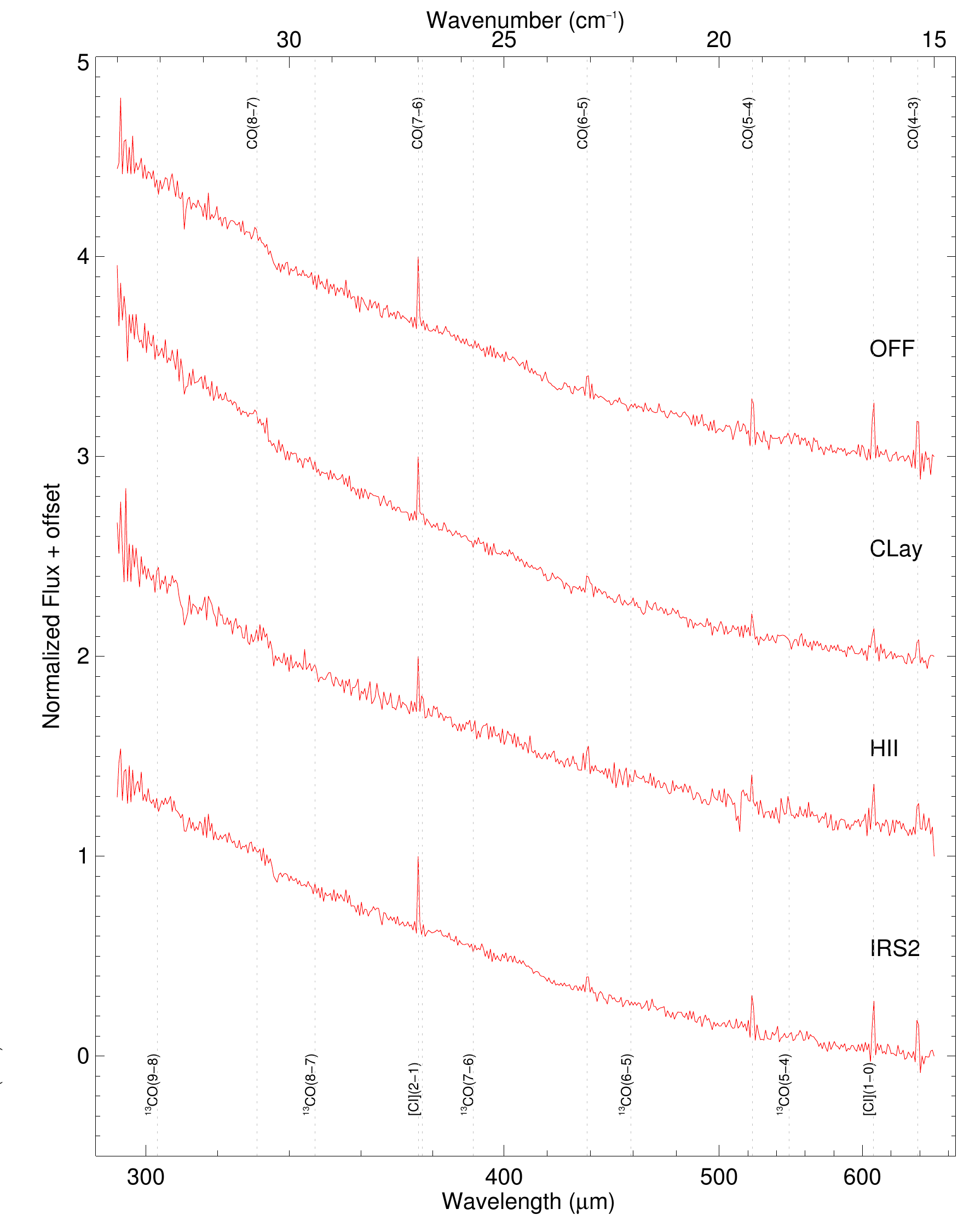}
	\caption{Same as Fig.~\ref{fig-spectra-lsw-h} but for the poorest pointings.}
	\label{fig-spectra-lsw-l}
\end{figure}

\begin{figure}
	\includegraphics[width=0.47\textwidth]{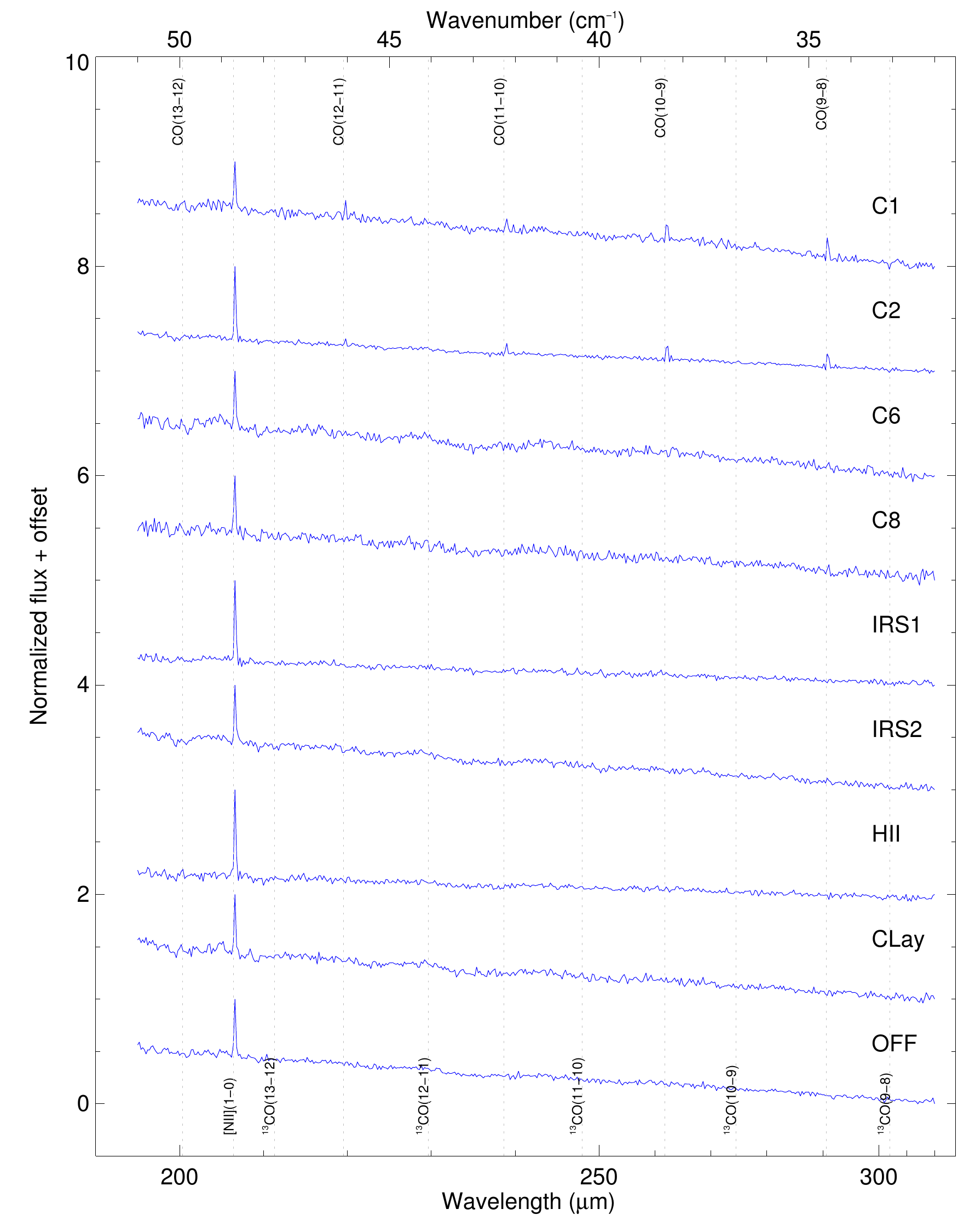}
	\caption{Same as Fig.~\ref{fig-spectra-lsw-h} but for all SPIRE-FTS SSW pointings.}
	\label{fig-spectra-ssw}
\end{figure}

OFF: the off position was chosen by looking at the \textit{Spitzer} mid-IR map of \mia. The PACS and SPIRE maps confirm that this is a low-emission region. However, after the analysis we find that this position has similar emission to other parts of the region (e.g., CLay, see Sec. \ref{sec-res}).

\begin{figure*}
	\includegraphics[width = 0.5\textwidth]{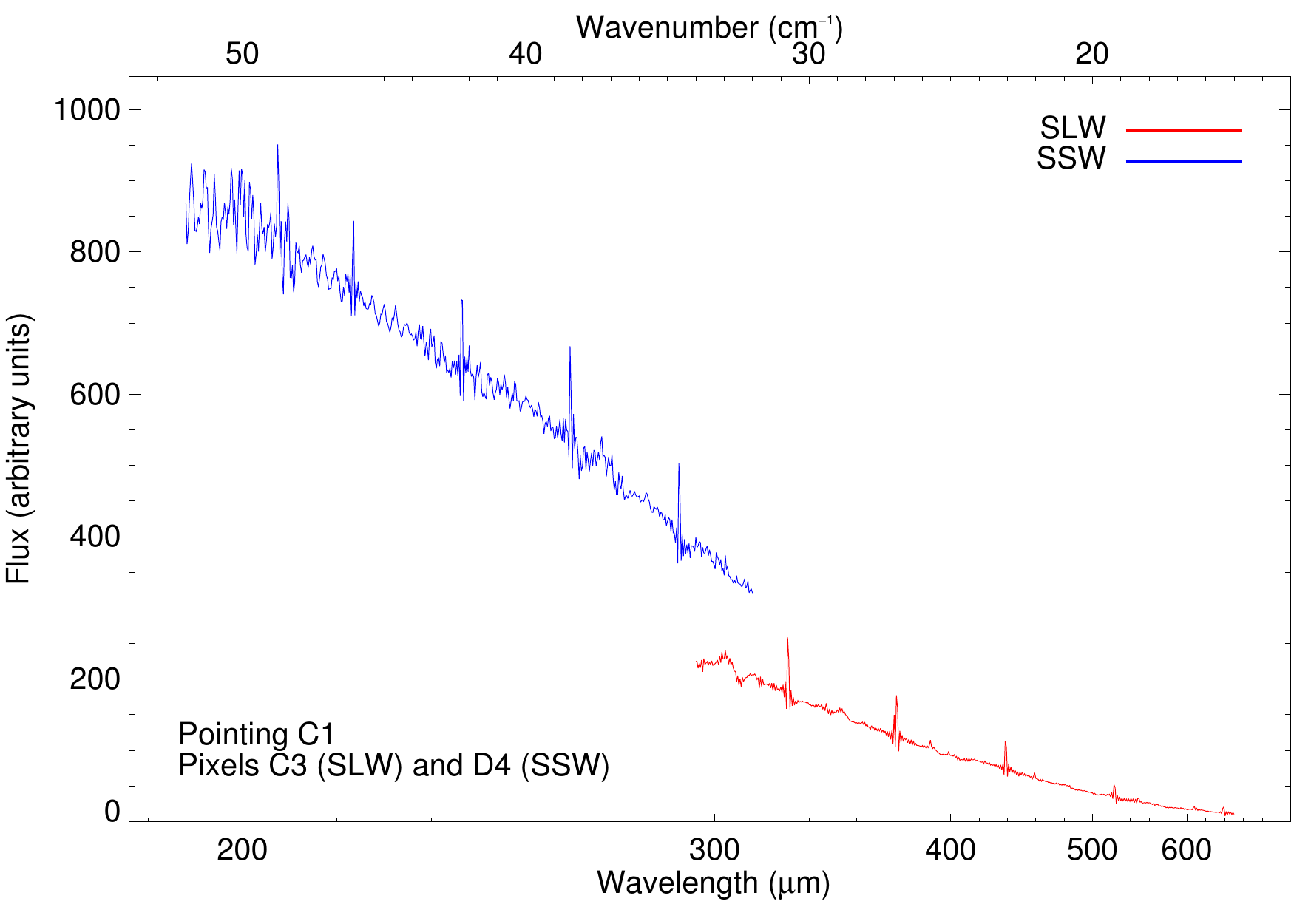}
	\includegraphics[width = 0.5\textwidth]{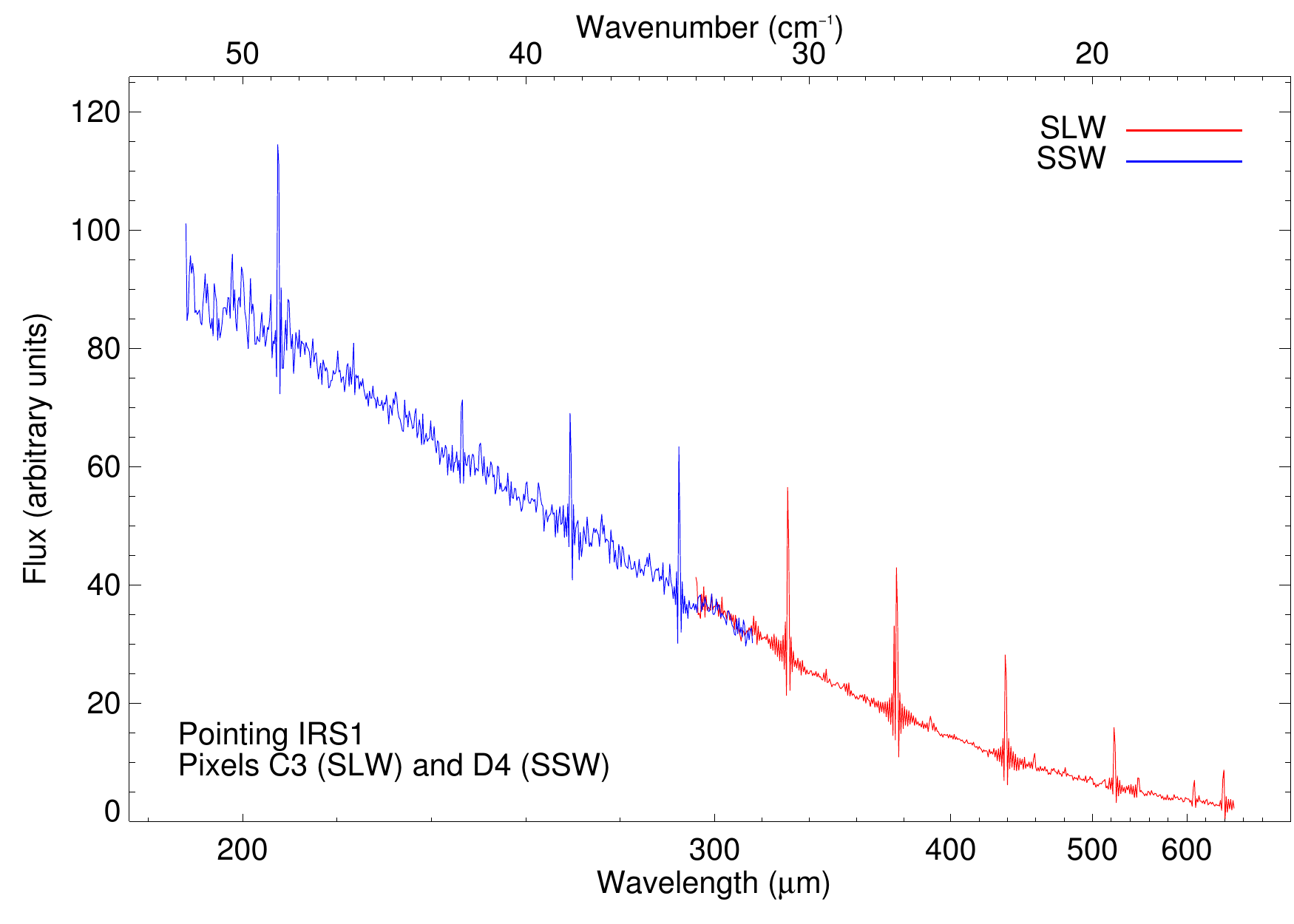}
	\caption{SPIRE-FTS SLW and SSW combined spectra for pointings C1 (left) and IRS1 (right), showing how in some cases the overlap between detector arrays is not perfect (see text).}
	\label{fig-gap}
\end{figure*}

\section{Results}
\label{sec-res}

\subsection{Measured spectra}
\label{sec-measured}

Each of the pixels of SLW and SSW arrays produces one spectrum, therefore we obtained $ 486 $ spectra. Owing to space constraints, we will only show a few selected spectra in this paper, in Figs.~\ref{fig-spectra-lsw-h}, \ref{fig-spectra-lsw-l} and \ref{fig-spectra-ssw}. These are the spectra at the central pixels of SLW and SSW for each pointing, which are the only cases where a spectrum on the whole wavelength range of the FTS at a given pointing can be obtained since the central pixels are the only ones whose centers are spatially coincident on the sky. For other pixels there certainly is a spatial overlap between the SLW and SSW (see Figs.~\ref{fig-pointings}~and~\ref{fig-detectors}), however the relative centers are shifted. 

The lines detected are marked in Figs.~\ref{fig-spectra-lsw-h} to \ref{fig-spectra-ssw}. These are the main lines observed in the obtained spectra, and the detections are discussed in the next section.

The overlap between the SLW and SSW portions of the spectra is not perfect. It can be seen in Fig. \ref{fig-gap} that in some cases there is a vertical offset between the two portions of the spectra.
This offset is due to the morphology of the regions mapped, and the complex properties of the SPIRE beam (see, e.g., \citealt{makiwa2013, spinoglio2012, fletcher2012}, and references therein). The flux calibration for the FTS observations uses a Relative Spectral Response Function based on the telescope model emission. By default it is assumed that the observed source is uniformly extended over the entire beam. Any source morphology (compact-like) would affect the calibration.
In our case, the result is that a relatively compact source (e.g., C1) will produce an apparently broken continuum, while for an extended source (e.g., IRS1) the detected signals are similar in both detectors.
New calibration tools have been developed for semi-extended sources (see \citealt{etxaluze2013, wu2013}). 
However, as we do not know the morphology of our source (which could vary with frequency), we are not able to use these tools. In Section \ref{sec:radex} we address the consequences this has on our results.

Because the FTS is a Fourier spectrograph, the line profiles have the characteristic \textit{sinc} function shape. This is more noticeable in the strong lines, for example the CO($6-5$) line at pointing IRS1, shown in Figs. \ref{fig-spectra-lsw-h} and \ref{fig-gap}.
The spectral resolution is $0.048\,{\rm cm}^{-1}$ throughout the FTS band \citep{swinyard2010}.

\begin{figure}
	\includegraphics[width=0.5\textwidth]{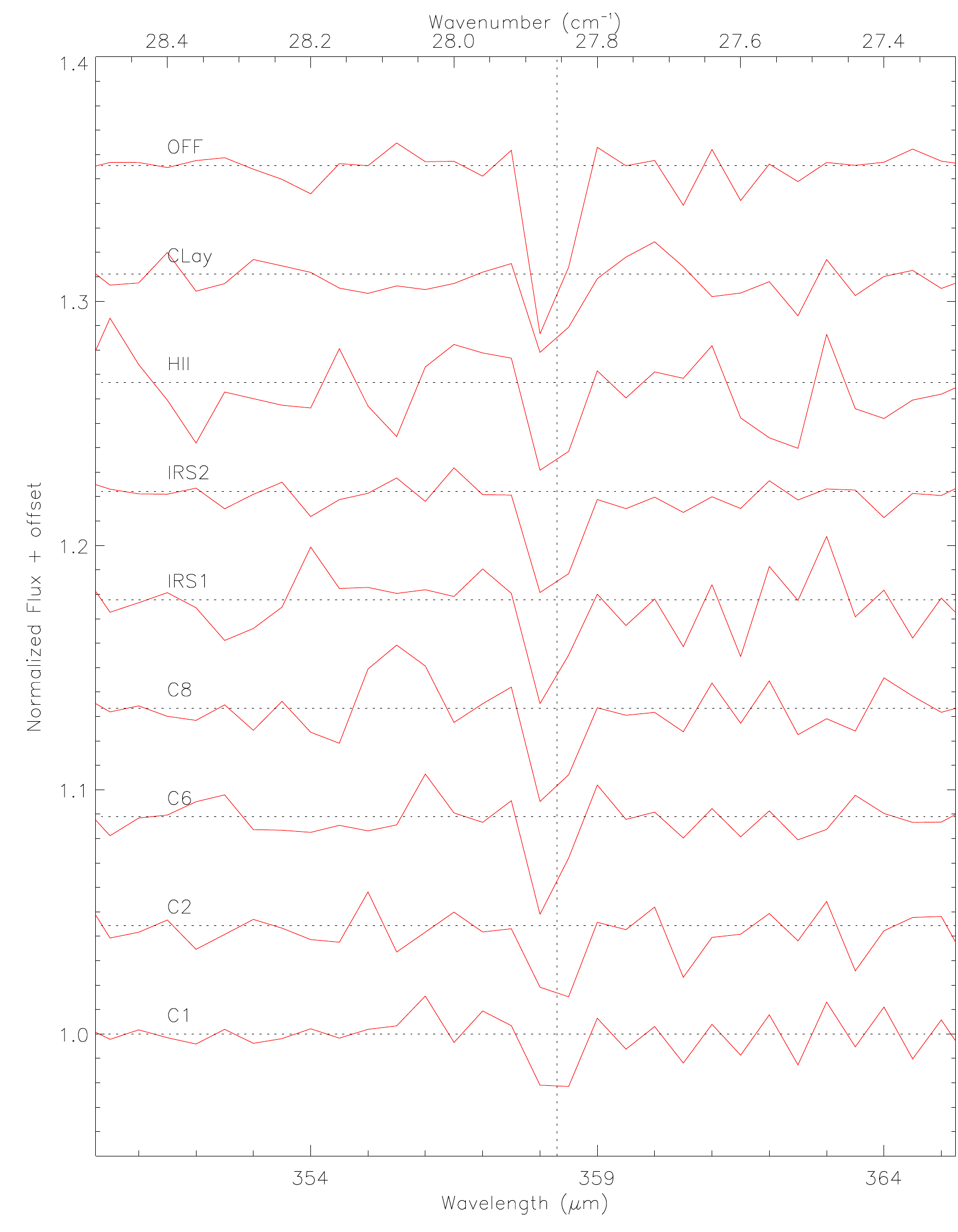}
	\caption{Zoom of the normalized CH$^{+}(1-0)$ absorption feature averaged over all pixels for each pointing.}
	\label{fig-ch+}
\end{figure}

\subsection{Detected line intensities}
\label{sec-lines}

\begin{table*}
	\renewcommand{\arraystretch}{1.2}  
    	\caption{Properties of the CO lines detected in emission in the central spectrum of each pointing.} 
    	\label{table-co}
     	\scriptsize
    	\centering
		\begin{tabular}{@{}l@{}@{}c@{ }|*{10}{r@{$\,\pm\,$}l}}
		\hline\hline
 &  & \multicolumn{20}{|c}{CO} \\
 &  & \multicolumn{2}{c}{4-3} & \multicolumn{2}{c}{5-4} & \multicolumn{2}{c}{6-5} & \multicolumn{2}{c}{7-6} & \multicolumn{2}{c}{8-7} & \multicolumn{2}{c}{9-8} & \multicolumn{2}{c}{10-9} & \multicolumn{2}{c}{11-10} & \multicolumn{2}{c}{12-11} & \multicolumn{2}{c}{13-12} \\
\hline
$\lambda$ & ($\mu$m) & \multicolumn{2}{c}{650.3} & \multicolumn{2}{c}{520.3} & \multicolumn{2}{c}{433.6} & \multicolumn{2}{c}{371.7} & \multicolumn{2}{c}{325.2} & \multicolumn{2}{c}{289.1} & \multicolumn{2}{c}{260.2} & \multicolumn{2}{c}{236.6} & \multicolumn{2}{c}{216.9} & \multicolumn{2}{c}{200.3} \\
E$_{up}$ & (K) & \multicolumn{2}{c}{55.3} & \multicolumn{2}{c}{83} & \multicolumn{2}{c}{116.2} & \multicolumn{2}{c}{154.9} & \multicolumn{2}{c}{199.1} & \multicolumn{2}{c}{248.9} & \multicolumn{2}{c}{304.2} & \multicolumn{2}{c}{365} & \multicolumn{2}{c}{431.3} & \multicolumn{2}{c}{503.1} \\
FWHM & ($''$) & \multicolumn{2}{c}{40.4} & \multicolumn{2}{c}{32.6} & \multicolumn{2}{c}{29.4} & \multicolumn{2}{c}{34.8} & \multicolumn{2}{c}{36.8} & \multicolumn{2}{c}{19.2} & \multicolumn{2}{c}{17.7} & \multicolumn{2}{c}{17.6} & \multicolumn{2}{c}{17} & \multicolumn{2}{c}{16.8} \\
\hline        																			
C1	&	\multirow{9}{*}{\rotatebox{90}{($10^{-3}\,$erg\,s$^{-1}$\,cm$^{-2}$\,sr$^{-1}$)}}	&	2.40	&	0.32	&	6.33	&	0.62	&	16.04	&	0.47	&	25.59	&	1.77	&	37.20	&	0.64	&	51.76	&	3.88	&	72.53	&	4.30	&	77.53	&	4.73	&	66.87	&	5.17	&	39.96	&	5.60	\\
C2	&		&	2.15	&	0.27	&	4.24	&	0.23	&	8.21	&	0.16	&	12.60	&	1.38	&	17.43	&	0.29	&	17.47	&	0.76	&	16.56	&	1.58	&	11.47	&	0.92	&	6.08	&	1.01	&	5.90	&	1.29	\\
C6	&		&	0.73	&	0.05	&	1.15	&	0.06	&	1.36	&	0.05	&	1.45	&	0.49	&	1.36	&	0.18	&	\multicolumn{2}{c}{\ldots}			&	\multicolumn{2}{c}{\ldots}			&	\multicolumn{2}{c}{\ldots}			&	\multicolumn{2}{c}{\ldots}			&	\multicolumn{2}{c}{\ldots}			\\
C8	&		&	0.92	&	0.25	&	1.55	&	0.27	&	2.18	&	0.17	&	2.02	&	0.42	&	1.37	&	0.40	&	\multicolumn{2}{c}{\ldots}			&	\multicolumn{2}{c}{\ldots}			&	\multicolumn{2}{c}{\ldots}			&	\multicolumn{2}{c}{\ldots}			&	\multicolumn{2}{c}{\ldots}			\\
IRS1	&		&	1.44	&	0.06	&	3.23	&	0.39	&	6.71	&	0.23	&	10.76	&	0.92	&	13.36	&	0.12	&	12.92	&	1.74	&	11.61	&	0.56	&	7.77	&	1.62	&	4.06	&	1.19	&	\multicolumn{2}{c}{\ldots}			\\
IRS2	&		&	0.43	&	0.06	&	0.50	&	0.10	&	0.29	&	0.07	&	\multicolumn{2}{c}{\ldots}			&	\multicolumn{2}{c}{\ldots}			&	\multicolumn{2}{c}{\ldots}			&	\multicolumn{2}{c}{\ldots}			&	\multicolumn{2}{c}{\ldots}			&	\multicolumn{2}{c}{\ldots}			&	\multicolumn{2}{c}{\ldots}			\\
HII	&		&	0.37	&	0.05	&	\multicolumn{2}{c}{\ldots}			&	\multicolumn{2}{c}{\ldots}			&	\multicolumn{2}{c}{\ldots}			&	\multicolumn{2}{c}{\ldots}			&	\multicolumn{2}{c}{\ldots}			&	\multicolumn{2}{c}{\ldots}			&	\multicolumn{2}{c}{\ldots}			&	\multicolumn{2}{c}{\ldots}			&	\multicolumn{2}{c}{\ldots}			\\
CLay	&		&	0.22	&	0.02	&	\multicolumn{2}{c}{\ldots}			&	\multicolumn{2}{c}{\ldots}			&	\multicolumn{2}{c}{\ldots}			&	\multicolumn{2}{c}{\ldots}			&	\multicolumn{2}{c}{\ldots}			&	\multicolumn{2}{c}{\ldots}			&	\multicolumn{2}{c}{\ldots}			&	\multicolumn{2}{c}{\ldots}			&	\multicolumn{2}{c}{\ldots}			\\
OFF	&		&	0.53	&	0.06	&	0.51	&	0.04	&	0.30	&	0.07	&	\multicolumn{2}{c}{\ldots}			&	\multicolumn{2}{c}{\ldots}			&	\multicolumn{2}{c}{\ldots}			&	\multicolumn{2}{c}{\ldots}			&	\multicolumn{2}{c}{\ldots}			&	\multicolumn{2}{c}{\ldots}			&	\multicolumn{2}{c}{\ldots}			\\

\hline\hline
\end{tabular}
\end{table*}

In the different spectra we detect the \co\ ladder from the $J=4\rightarrow3$ to the $J=13\rightarrow12$ transitions; the \tco\ ladder between the $J=5\rightarrow4$ and the $J=8\rightarrow7$ transitions; the [CI] $^{3}P_{1}-^{3}P_{0}$ and $^{3}P_{2}-^{3}P_{1}$ transitions ([CI]($1-0$) and [CI]($2-1$) from now on); and the [NII] $^{3}P_{1}-^{3}P_{0}$ transition ([NII]($1-0$) from now on). In Tables \ref{table-co}, \ref{table-13co}, and \ref{table-lines} are the details of the lines detected in emission on each of the central spectra, shown in Figs. \ref{fig-spectra-lsw-h}, \ref{fig-spectra-lsw-l} and \ref{fig-spectra-ssw}.

In the pointings towards a condensation (C1, C2, C6, C8, IRS1, and IRS2) or the PDR (CLay) of \mia, the number of lines detected on each pixel correlate with the position of said pixel in the sky in such a way that the pixels towards the PDR material or the condensation show richer spectra than the pixels ``off'' the PDR or the condensation. This means that for example in pointing C1 pixels C1, D1, and E1 show fewer lines than the others.
In the other pointings, towards diffuse gas (HII and OFF), the lines detected are mostly the same for all pixels of a given pointing. The only line that is detected in every pixel of every pointing is the [NII]($1-0$) transition. The [CI] transitions are present in almost all the pixels. This is also true for the first three CO transitions detected, except in pointings HII and \pdr, where only the lowest CO transition detected, i.e., $J=4\rightarrow3$, is present. The $^{13}$CO lines are detected only in pointings C1, C2, and IRS1. Those pointings are also the only ones with detections of the higher CO transitions.

The CH$^{+}(1-0)$ line has been also detected, but in all the cases it appears in absorption. 
This CH$^{+}$ transition is seen in emission in most of the Orion Bar \citep{naylor2010} while it is seen in absorption in two ultracompact \hii\ (UC\hii) regions located near the Galactic plane \citep{kirk2010}. Therefore we can expect that the corresponding lines seen in the direction of \mia\ should be the result of both emission and absorption mechanisms.
Because of the low spectral resolution of the FTS these lines are not spectrally resolved and we cannot obtain their equivalent width, thus no quantitative results can be derived from the FTS spectra. Nevertheless, we can see from Fig.~\ref{fig-ch+} that the largest absorption seems to be in the direction of the OFF position, while the pointings towards the PDR of \mia\ show the weakest signatures. All this suggests that CH$^{+}(1-0)$ would be tracing diffuse gas along the line of sight, while \mia\ is in fact emitting in this line, dampening the absorption in the pointings towards the \hii\ region (see, e.g., \citealt{falgarone2010,nagy2013}).

\begin{table}
	\renewcommand{\arraystretch}{1.2}  
    	\caption{Same as Table~\ref{table-co} but for \tco} 
    	\label{table-13co}
     	\small
    	\centering
		\begin{tabular}{@{}l@{}@{}c@{ }|*{4}{r@{$\,\pm\,$}l}}
		\hline\hline
& & \multicolumn{8}{|c}{$^{13}$CO} \\
& & \multicolumn{2}{c}{5-4} & \multicolumn{2}{c}{6-5} & \multicolumn{2}{c}{7-6} & \multicolumn{2}{c}{8-7} \\ 
\hline																		
$\lambda$ & ($\mu$m) & \multicolumn{2}{c}{544.2} & \multicolumn{2}{c}{453.5} & \multicolumn{2}{c}{388.7} & \multicolumn{2}{c}{340.2} \\ 
E$_{up}$ & (K) & \multicolumn{2}{c}{79.3} &\multicolumn{2}{c}{111.1} & \multicolumn{2}{c}{148.1} & \multicolumn{2}{c}{190.4} \\ 
FWHM & ($''$) & \multicolumn{2}{c}{32.9} & \multicolumn{2}{c}{30} & \multicolumn{2}{c}{34} & \multicolumn{2}{c}{36.1} \\ 
\hline																	
C1	&	\multirow{9}{*}{\rotatebox{90}{($10^{-3}\,$erg\,s$^{-1}$\,cm$^{-2}$\,sr$^{-1}$)}}	&	1.48	&	0.37	&	2.35	&	0.45	&	3.74	&	0.52	&	3.97	&	0.60	\\
C2	&		&	0.86	&	0.09	&	1.23	&	0.31	&	1.54	&	0.13	&	0.94	&	0.31	\\
C6	&		&	\multicolumn{2}{c}{\ldots}			&	\multicolumn{2}{c}{\ldots}			&	\multicolumn{2}{c}{\ldots}			&	\multicolumn{2}{c}{\ldots}			\\
C8	&		&	\multicolumn{2}{c}{\ldots}			&	\multicolumn{2}{c}{\ldots}			&	\multicolumn{2}{c}{\ldots}			&	\multicolumn{2}{c}{\ldots}			\\
IRS1	&		&	0.73	&	0.16	&	0.71	&	0.28	&	0.78	&	0.10	&	0.65	&	0.11	\\
IRS2	&		&	\multicolumn{2}{c}{\ldots}			&	\multicolumn{2}{c}{\ldots}			&	\multicolumn{2}{c}{\ldots}			&	\multicolumn{2}{c}{\ldots}			\\
HII	&		&	\multicolumn{2}{c}{\ldots}			&	\multicolumn{2}{c}{\ldots}			&	\multicolumn{2}{c}{\ldots}			&	\multicolumn{2}{c}{\ldots}			\\
CLay	&		&	\multicolumn{2}{c}{\ldots}			&	\multicolumn{2}{c}{\ldots}			&	\multicolumn{2}{c}{\ldots}			&	\multicolumn{2}{c}{\ldots}			\\
OFF	&		&	\multicolumn{2}{c}{\ldots}			&	\multicolumn{2}{c}{\ldots}			&	\multicolumn{2}{c}{\ldots}			&	\multicolumn{2}{c}{\ldots}			\\
\hline\hline
\end{tabular}
\end{table}

\begin{table}
	\renewcommand{\arraystretch}{1.2}  
    	\caption{Same as Table~\ref{table-co} but for \textbf{atomic} species} 
    	\label{table-lines}
    	\centering
		\begin{tabular}{@{}l@{}@{}c@{ }|*{2}{r@{$\,\pm\,$}l}|r@{$\,\pm\,$}l}
		\hline\hline
& & \multicolumn{4}{|c|}{[CI]} & \multicolumn{2}{c}{[NII]} \\
 & & \multicolumn{2}{c}{1-0} & \multicolumn{2}{c|}{2-1} & \multicolumn{2}{c}{1-0} \\
\hline
$\lambda$ & ($\mu$m) & \multicolumn{2}{c}{609.1} & \multicolumn{2}{c|}{370.4} & \multicolumn{2}{c}{205.2} \\
E$_{up}$ & (K) & \multicolumn{2}{c}{23.6} & \multicolumn{2}{c|}{62.5} & \multicolumn{2}{c}{70.1} \\
FWHM & ($''$) & \multicolumn{2}{c}{37.2} & \multicolumn{2}{c|}{34.8} & \multicolumn{2}{c}{16.9} \\
\hline													
C1	&	\multirow{9}{*}{\rotatebox{90}{($10^{-3}\,$erg\,s$^{-1}$\,cm$^{-2}$\,sr$^{-1}$)}}	&	1.12	&	0.34	&	6.80	&	0.57	&	84.99	&	5.46	\\
C2	&		&	1.01	&	0.28	&	5.50	&	0.20	&	46.04	&	3.87	\\
C6	&		&	0.61	&	0.06	&	1.92	&	0.06	&	10.02	&	0.39	\\
C8	&		&	0.67	&	0.19	&	1.66	&	0.11	&	15.06	&	1.02	\\
IRS1	&		&	0.95	&	0.14	&	3.72	&	0.26	&	26.33	&	0.72	\\
IRS2	&		&	0.53	&	0.06	&	1.18	&	0.05	&	13.27	&	1.45	\\
HII	&		&	0.48	&	0.05	&	0.88	&	0.11	&	34.39	&	1.24	\\
CLay	&		&	0.29	&	0.03	&	0.87	&	0.09	&	13.13	&	1.20	\\
OFF	&		&	0.53	&	0.07	&	0.98	&	0.08	&	19.96	&	0.34	\\
\hline\hline												
	\end{tabular}
\end{table}

\begin{figure*}[t]
\includegraphics[width=0.99\textwidth]{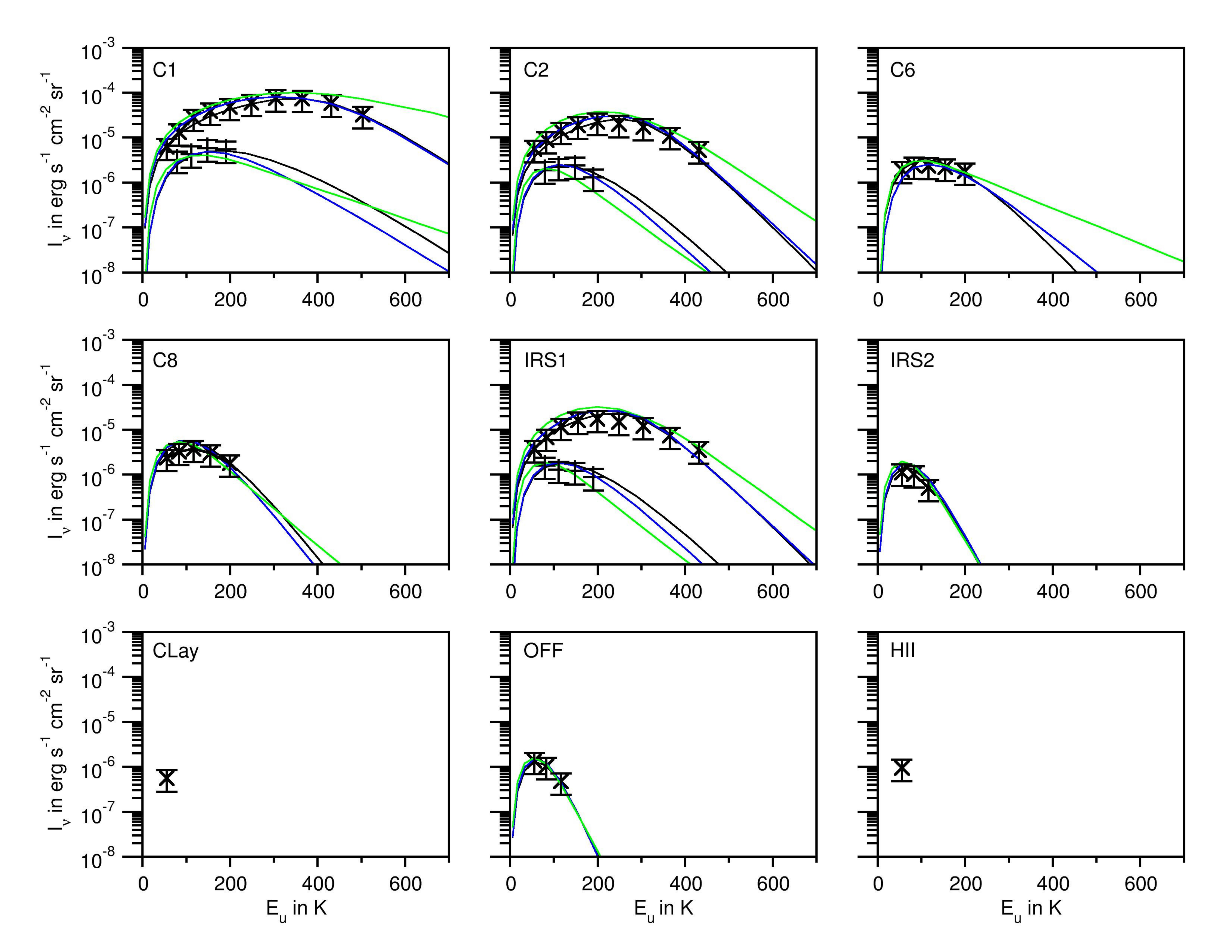}
\caption[]{Fit of integrated line intensities of the observed $^{12}$CO and $^{13}$CO lines (symbols) with RADEX for three densities: $n=10^4$ (green lines), $n=10^5$ (blue lines) and $n=10^6$~cm$^{-3}$ (black lines). The resulting temperatures and CO column densities are shown in Table~\ref{tab:radex}.}
\label{fig:denprof}
\end{figure*}

\begin{center}
\begin{figure*}[t]
\includegraphics[width=0.99\textwidth]{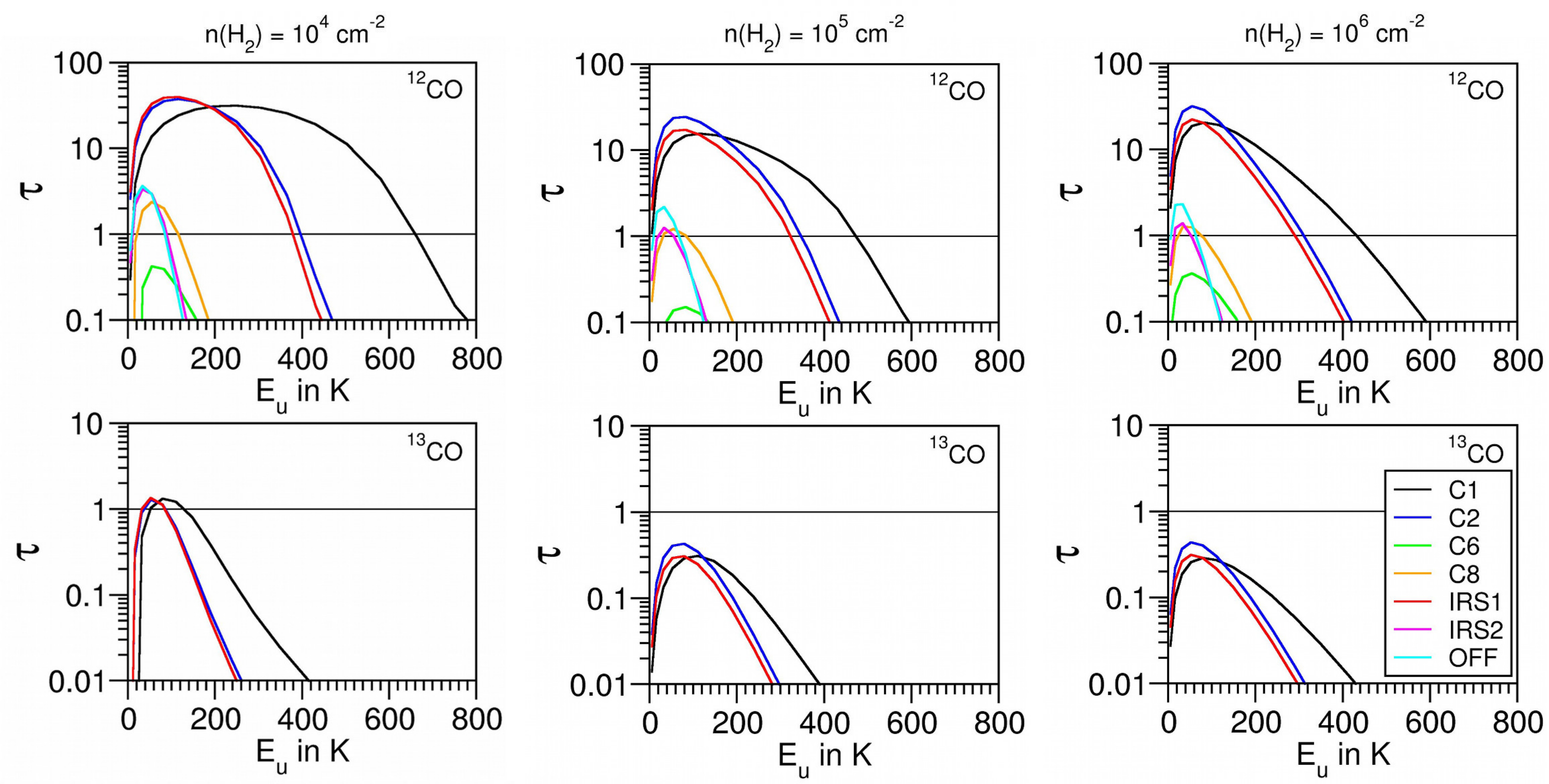}
\caption[]{Optical depth $\tau$ determined with RADEX for $^{12}$CO (top) and $^{13}$CO (bottom) for n$_{\rm H_2}=10^4\,{\rm cm^{-2}}$ (left), n$_{\rm H_2}=10^5\,{\rm cm^{-2}}$ (middle) and n$_{\rm H_2}=10^6\,{\rm cm^{-2}}$ (right).}
\label{fig:tau}
\end{figure*}
\end{center}

\subsection{Determination of physical properties using RADEX}
\label{sec:radex}

In Fig. \ref{fig:denprof} we present the integrated line intensities of the different positions for $^{12}$CO (crosses) and $^{13}$CO (plus signs).

Comparing the curves of the integrated line intensities, four groups can be distinguished:
\begin{enumerate}
\item C1, C2 and IRS1 show high-excited $^{12}$CO and $^{13}$CO lines and a small $^{12}$CO/$^{13}$CO ratio for low-excited lines,
\item C6 and C8 show no $^{13}$CO lines, no high-excited $^{12}$CO lines and the low-excited $^{12}$CO lines have a moderately high intensity,
\item IRS2 and OFF have no $^{13}$CO lines, no high-excited $^ {12}$CO lines and the low-excited $^{12}$CO lines have a rather low intensity, and
\item CLay and HII show only one, the $^{12}$CO($4-3$) line.
\end{enumerate}

In order to asses the kinetic temperature, density and column density in which the CO lines arise in the different positions, we fit the integrated line intensities of the observed $^{12}$CO and $^{13}$CO lines with the non-LTE and (local) radiative transfer code RADEX\footnote{\url{http://www.sron.rug.nl/~vdtak/radex/index.shtml}.}.
We use a grid of input parameters, where the gas density varies between 10$^4$\,cm$^{-3}$ and 10$^6$\,cm$^{-3}$, the kinetic temperatures between 10\,K and 500\,K, CO column densities between 10$^{15}$\,cm$^{-2}$ to 10$^{19}$\,cm$^{-2}$ and the beam filling factor $\eta$ between 0.01 and 1.
We use the new set of collisional rate coefficients calculated by \citet{yang2010} including energy levels up to J = 40 for temperatures ranging from 2\,K to 3000\,K. 
A standard carbon isotopic ratio $^{12}$C/$^{13}$C of 70 \citep{wilson1999} is assumed.
We use the cosmic microwave background radiation at 2.73\,K.

The effect introduced by the default assumption in the calibration is reflected in a small gap in Fig.~\ref{fig:denprof} between the intensities of the CO lines at the edge of the SSW and SLW receivers. However this gap is smaller than the $ 30\% $ error bars assumed for the calculations.
Therefore, the inability to use the semi-extended calibration tools mentioned in Section~\ref{sec-measured} has no measurable effect on the temperatures and densities we derive.

We fit the slope of the cooling curve of the CO lines with different combinations of kinetic temperature and gas density, which reflects a degeneracy between these two parameters.
For a given combination of kinetic temperature and gas density, we obtain the beam-averaged column density by fitting the line ratio $^{12}$CO/$^{13}$CO which is sensitive to optical effects\footnote{In the cases where $^{13}$CO is not detected, we are only able to give a lower limit of the column density, assuming a filling factor of one.}.
The optical depth at the line center depends on the ratio of the column density to line width.
For reasons of simplicity, we take a constant width for all the lines measured in the FTS cubes. We assume $\Delta \rm{v} = 2~{\rm km~s^{-1}}$, since for the range and the physical conditions considered we find that the RADEX results do not vary significantly for $\Delta \rm{v} = 0.5 - 2~{\rm km~s^{-1}}$. Those values are in agreement with HiFi observations towards classical PDRs (e.g., NGC~7023 \citealt{ossenkopf2013}).
The beam filling factor was obtained by fitting the mean absolute line intensities.

The results for three density values, n$_{\rm H_2}=10^4$, n$_{\rm H_2}=10^5$ and n$_{\rm H_2}=10^6$\,cm$^{-3}$, are presented in Figs.~\ref{fig:denprof} and \ref{fig:tau}, and are summarized in Table~\ref{tab:radex}.
For each set of parameters, we derive the length of the emission layer along the line of sight: $l\sim N_{\rm H_2}/n_{\rm H_2}$. 
We convert the CO into H$_2$ column densities taking a standard relative $^{12}$CO abundance to H$_2$ of $8\times10^{-5}$ in PDRs \citep{johnstone2003}.

\begin{table*}[t]
	\renewcommand{\arraystretch}{1.2}
	\centering
	\caption{The results with RADEX for the different positions.}
	\begin{tabular}{l|cccc|cccc|cccc}
	\hline\hline
	 \T \B& \multicolumn{4}{c|}{n$_{\rm H_2}=10^6{\rm cm^{-3}}$} & \multicolumn{4}{c|}{n$_{\rm H_2}=10^5{\rm cm^{-3}}$} & \multicolumn{4}{c}{n$_{\rm H_2}=10^4{\rm cm^{-3}}$} \\
 \cline{2-13}
 \T \B & T & ${\rm N_{\rm CO}}$ & l & $\eta$ & T & ${\rm N_{\rm CO}}$ & l & $\eta$ & T & ${\rm N_{\rm CO}}$ & l & $\eta$ \\
 \B & [K] & ($ 10^{16}\,$cm$^{-2}$) & ($ 10^{-3}\,$pc) & & (K) & ($ 10^{16}\,$cm$^{-2}$) & ($ 10^{-3}\,$pc) & & (K) & ($ 10^{16}\,$cm$^{-2}$) & ($ 10^{-3}\,$pc) & \\ 
\hline
C1 \T & 70 & $100$ & $4$ & 0.5 & 100 & $100$ & 40 & 0.5 & 250 & $300$ & 1200 & 0.3 \\
C2 & 45 & $100$ & $4$ & 0.5 & 60 & $100$ & 40 & 0.5 & 100 & $200$ & 800 & 0.6 \\
C6 & 40 & $>1$ & $>0.05$ & \ldots & 70 & $>0.5$ & $>0.2$ & \ldots & 200 & $>1$ & $>4$ & \ldots \\
C8 & 33 & $>3$ & $>0.1$ & \ldots & 42 & $>3$ & $>1$ & \ldots & 80 & $>5$ & $>20$ & \ldots \\
IRS1 & 45 & $70$ & $3$ & 0.5 & 60 & $70$ & 30 & 0.5 & 90 & $200$ & $800$ & 0.5 \\
IRS2 & 20 & $>2$ & $>0.08$ & \ldots & 25 & $>2$ & $>0.8$ & \ldots & 35 & $>5$ & $>20$ & \ldots \\
OFF & 17 & $>3$ & $>0.1$ & \ldots & 20 & $>3$ & $>1$ & \ldots & 30 & $>5$ & $>20$ & \ldots \\
\hline\hline
\end{tabular}
\label{tab:radex}
\end{table*}

Positions C1, C2, and IRS1 are rather similar concerning the gas line analysis.
We obtain relatively high column densities, $N_{\rm CO}$, of around $1 \times 10^{18}$ ${\rm cm^{-2}}$ of warm and dense gas for C1, C2 and IRS1. 
The CO temperatures are rather similar in these three position while pointing C1 might have slightly higher temperatures.
It is rather difficult to fix the density, since with all three assumed density values of n$_{\rm H_2}=10^4$, n$_{\rm H_2}=10^5$ and n$_{\rm H_2}=10^6$~cm$^{-3}$ we are able to fit the observations. The differences lie in the obtained length along the line of sight for these three assumptions. Where for n$_{\rm H_2}=10^4\,{\rm cm}^{-3}$ the length is rather large, of around $1\,$pc, for n$_{\rm H_2}=10^6\,{\rm cm}^{-3}$ the length is rather small, of around $0.003\,$ to $ 0.004\,$pc. 
The observed projected width and extension of the CO emission lines $(J\ge9)$ deduced from the detection in the detectors are $\le0.1\,$pc and $\sim0.5\,$pc, respectively. This roughly follows the emission of the dust as seen in Fig.~\ref{fig-pointings}. The calculated length along the line of sight from the RADEX fit is about two times larger than the extension for n$_{\rm H_2}=10^4\,{\rm cm}^{-3}$, 3 times smaller to the width for n$_{\rm H_2}=10^5\,{\rm cm}^{-3}$ and $25-30$ times smaller than the upper width limit for n$_{\rm H_2}=10^6\,{\rm cm}^{-3}$.
The large length along the line of sight for n$_{\rm H_2}=10^4\,{\rm cm}^{-3}$ exclude lower gas densities.
The small length along the line of sight derived for n$_{\rm H_2}=10^6\,{\rm cm}^{-3}$ could  be the result of clumps.
Regarding the comparison of projected extension, $0.1-0.5\,$pc, and obtained length from the model calculations, we would therefore tend to a density between n$_{\rm H_2}=10^4\,{\rm cm}^{-3}$ and n$_{\rm H_2}=10^5\,{\rm cm}^{-3}$.

Regarding the optical depth, we see from Fig.~\ref{fig:tau} that in pointing C1 all the detected $^{12}$CO lines are optically thick (E$_{up}\sim 55 - 500\,$K), while for pointings C2 and IRS1 the uppermost transitions detected have optical depths $ \tau < 1 $. The $ ^{13} $CO lines are mostly optically thin for all positions.

For the positions C6, C8, IRS2 and OFF, no $^{13}$CO and no high-excited $^{12}$CO lines are detected, which suggest that the column density of warm gas is rather small. 
The obtained lower limit on the CO column densities in these four regions are similar around  $1 - 5 \times 10^{16}\,{\rm cm^{-2}}$.
The density and temperature cannot be unambiguously determined.

In position C6 a low density of n$_{\rm H_2}=10^4$\,cm$^{-3}$ results in a very high gas temperature. 
It might be more likely to find lower gas temperatures which would lead to slightly larger densities.

For the positions C8, IRS2 and OFF, it is not possible to make constraint on the density. For all assumed densities the gas temperatures are smaller than in the other positions.

For the position OFF we find the smallest temperatures, which is not that surprising since the position is at a far distance from the star and no YSO is assumed in its vicinity. Whether this position is located in a dense molecular cloud or in a diffuse environment cannot unambiguously be determined.

For position CLay and HII only the $^{12}$CO J=4-3 line is detected, therefore we are not able to carry out a RADEX fit. 
The lack of observed higher-excited $^{12}$CO indicates a low temperature and/or a low density. 
It can further be assumed that the small integrated line intensity of this one line is the result of a rather small column density.

In summary, we get a relatively good assumption of the physical properties for positions C1, C2 and IRS1, while for the other positions we can only narrow down the physical properties.

\begin{figure*}
	\includegraphics[width=0.5\textwidth]{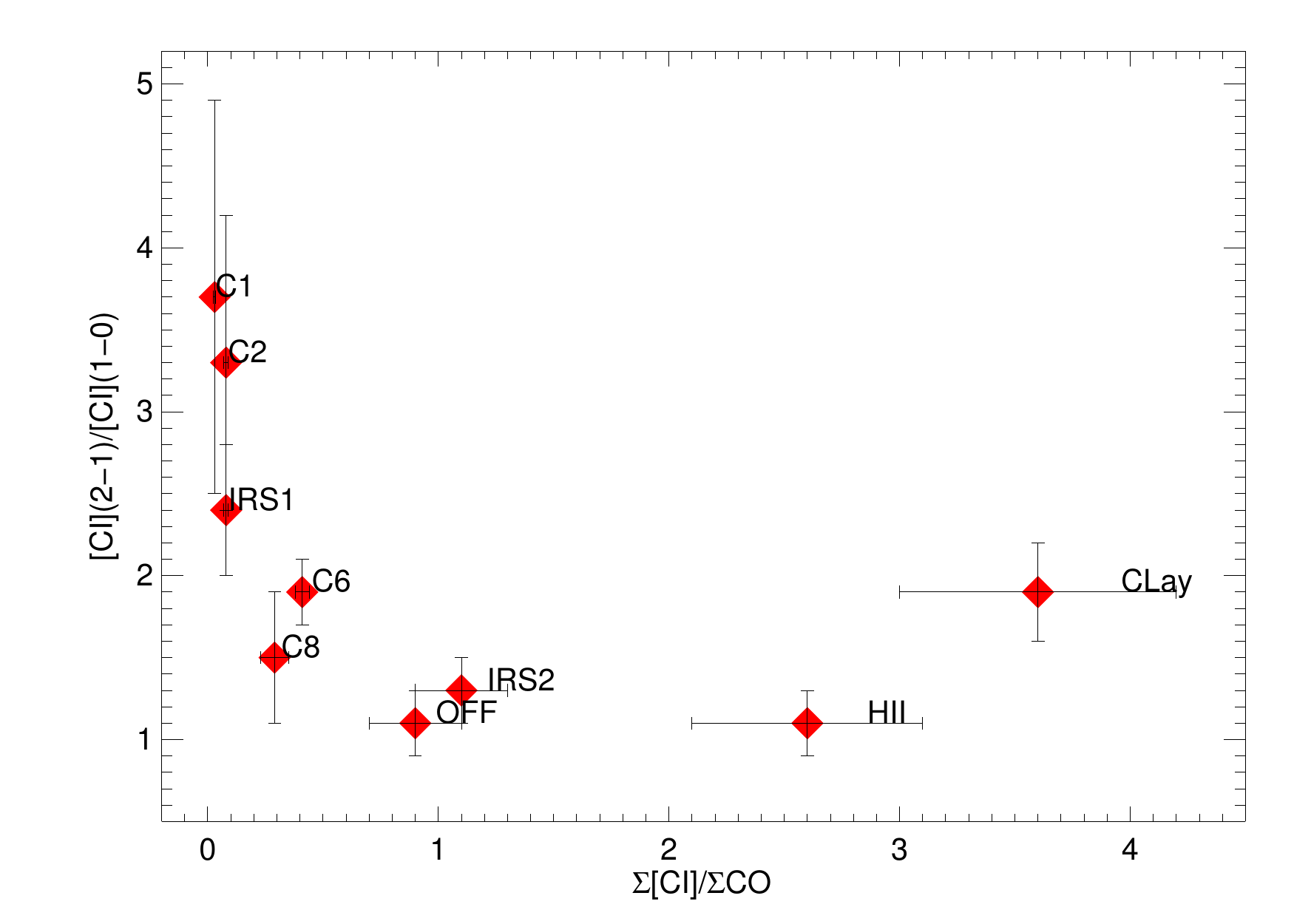}
	\includegraphics[width=0.5\textwidth]{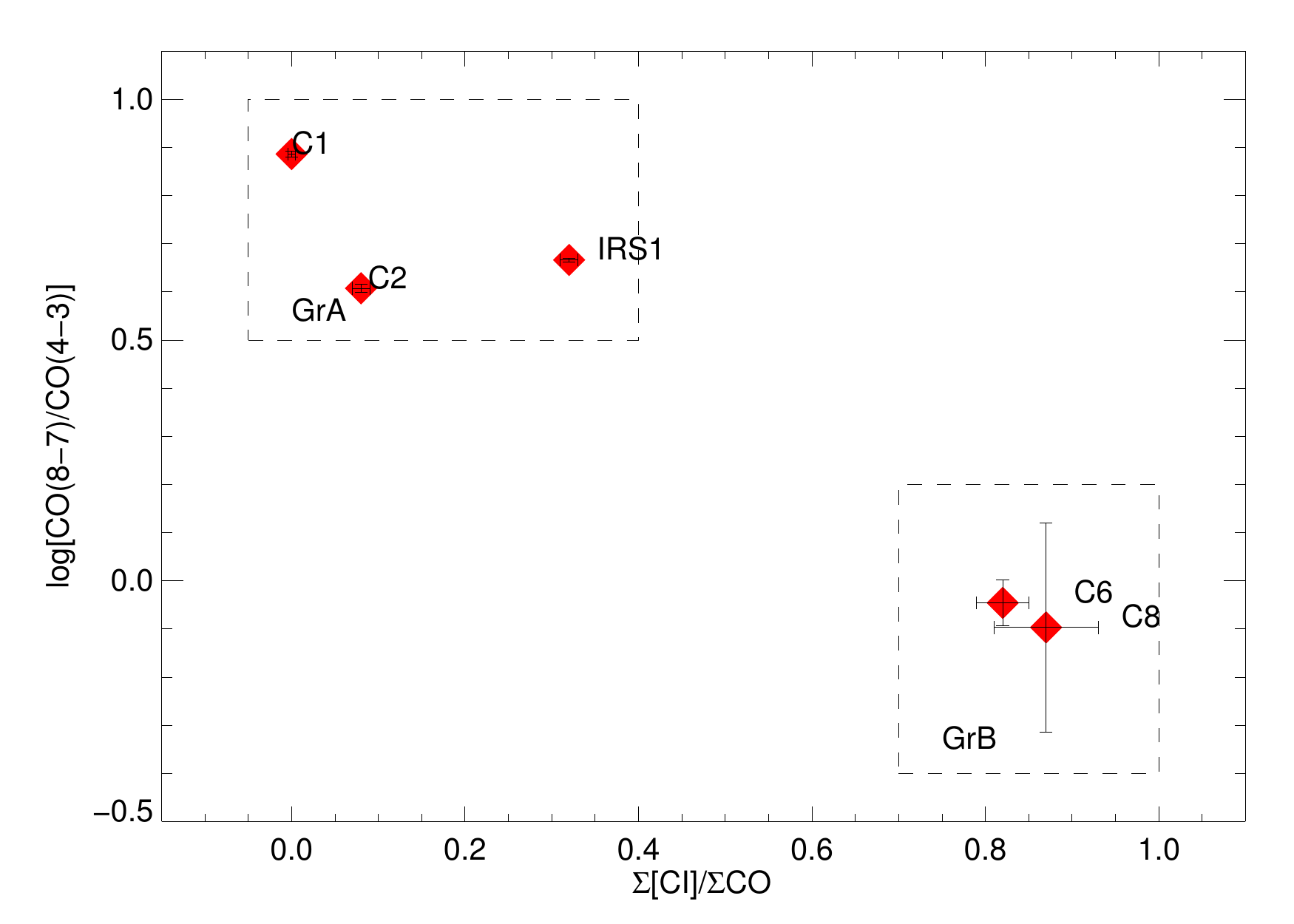}
	\caption{Line intensity ratios for the central pixel (C3) or each pointing.}
	\label{fig-ratios}
\end{figure*}

\begin{table}
	\renewcommand{\arraystretch}{1.2}
	\centering
	\caption{Line intensity ratios for the central pixel (C3) of each pointing.}
	\label{table-ratios}
	\begin{tabular}{l|ccc}
		\hline\hline
		Pointing & \cir & \rco & \cicotr \\
		\hline
C1	&	3.7	$\pm$	1.2	&	7.7	$\pm$	0.1	&	0.030	$\pm$	0.004	\\
C2	&	3.3	$\pm$	0.9	&	4.05	$\pm$	0.08	&	0.08	$\pm$	0.01	\\
C6	&	1.9	$\pm$	0.2	&	0.9	$\pm$	0.1	&	0.41	$\pm$	0.03	\\
C8	&	1.5	$\pm$	0.4	&	0.8	$\pm$	0.4	&	0.29	$\pm$	0.06	\\
IRS1	&	2.4	$\pm$	0.4	&	4.64	$\pm$	0.04	&	0.08	$\pm$	0.01	\\
IRS2	&	1.3	$\pm$	0.2	&	\ldots	&	1.1	$\pm$	0.2	\\
HII	&	1.1	$\pm$	0.2	&	\ldots	&	2.6	$\pm$	0.5	\\
CLay	&	1.9	$\pm$	0.3	&	\ldots	&	3.6	$\pm$	0.6	\\
OFF	&	1.1	$\pm$	0.2	&	\dots	&	0.9	$\pm$	0.2	\\
		\hline\hline
	\end{tabular}
\end{table}

\subsection{Line ratios}

In Table~\ref{table-ratios} we give the line ratios \mbox{\cicotr\,=\,$\sum$[CI]/$\sum$CO}, \mbox{\cir\,=\,[CI]($ 2-1 $)/[CI]($ 1-0 $)}, and \mbox{\rco\,=\,CO($ 8-7 $)/CO($ 4-3 $)} for the central pixel of each pointing. In the right panel of Fig.~\ref{fig-ratios} we see how the pointings can be separated into two groups according to the \rco\ and \cicotr\ ratios, labeled \gra\ (C1, C2, IRS1) and \grb\ (C6, C8) in the figure. A third group \grc, not shown in the figure, can be defined containing those pointings that do not have CO($ 8-7 $) emission (IRS2, \pdr, HII, OFF)

For \gra\ the \rco\ ratio is highest, signaling a high density and/or temperature, while the \cicotr\ ratio is the lowest, suggesting that the cooling by CO lines is more important than by [CI] emission and therefore that the density is high.

For \grb, the \rco\ ratio is low and the \cicotr\ ratio is high. The high \cicotr\ ratio indicates a low density and the low \rco\ ratio indicates a low density and/or temperature.

The pointings in \grc\ have the lowest \rco\ ratio, since the CO($ 8-7 $) line is undetectable, and the highest \cicotr\ ratio. This shows that cooling by [CI] emission is larger than by CO, and much more important than in the other two groups. This medium should be therefore less dense compared to \grb.
This behavior is largely consistent with the temperatures obtained with RADEX in Sec.~\ref{sec:radex}.

Taking into account the position of each pointing (see Fig.~\ref{fig-pointings}), we see that the groups can also be related to a distinction between regions of \mia; group \gra\ are the YSOs located in the PDR, \grb\ are the dust condensations situated not in the PDR but next to it, and in \grc\ are the pointings targeting regions of \mia\ with diffuse gas. 
As seen before in section \ref{sec:radex}, pointing IRS2 can be associated with the diffuse gas regions (\grc), despite being targeting a known protostellar source.

In the optically thin limit and assuming LTE, the ratio of the velocity integrated emission of [CI]($ 2-1 $) and [CI]($ 1-0 $) is a sensitive function of the excitation temperature, on the form $T_{ex} = 38.3\,{\rm K}/ \ln[2.11/\cir]$ (e.g., \citealt{kramer2004}). Using this equation, we calculated the temperatures for the central pixels of each pointing, shown in column 3 of Table~\ref{table-citemp}.
These values are different as the ones derived with RADEX.
The larger differences are for the pointings targeting regions off the PDR (IRS2 and OFF), which lead us to suggest that a large fraction of the [CI] emission is likely originating from the warmer surface layers.
This is also hinted by the [CI] temperatures, since all of them are similar independent of the position in the cloud.

\begin{table}
	\renewcommand{\arraystretch}{1.2}
	\centering
	\caption{Velocity integrated [CI]($ 2-1 $)/[CI]($ 1-0 $) ratios and temperatures associated with it for pixel C3 of each pointing.}
	\label{table-citemp}
	\begin{tabular}{l|c|c}
		\hline\hline
		Pointing & Ratio & T$_{ex}$ (K) \\
		\hline
C1	&	1.6	$\pm$	0.5	&	138	$\pm$	12	\\
C2	&	1.6	$\pm$	0.4	&	131	$\pm$	11	\\
C6	&	1.5	$\pm$	0.2	&	107	$\pm$	4	\\
C8	&	1.4	$\pm$	0.4	&	98	$\pm$	11	\\
IRS1	&	1.5	$\pm$	0.3	&	114	$\pm$	6	\\
IRS2	&	1.4	$\pm$	0.2	&	96	$\pm$	4	\\
HII	&	1.4	$\pm$	0.2	&	91	$\pm$	6	\\
CLay	&	1.5	$\pm$	0.2	&	109	$\pm$	6	\\
OFF	&	1.4	$\pm$	0.2	&	91	$\pm$	6	\\
		\hline\hline
	\end{tabular}
\end{table}

\subsection{PDR model}
\label{sec-model}

In order to obtain a first approach to the gas density and the UV radiation field we applied the Meudon PDR model, developed by \citet{lepetit2006}.
This is a 1-D radiative transfer model, which consists of a plane-parallel gas and dust slab of a given depth, illuminated on one or both sides by an ultraviolet (UV) radiation field and observed in the face-on direction. The slab depth is measured by its visual extinction $\av$, in magnitudes. The UV field is in units of the local interstellar value of $5.6\times10^{-14}\,{\rm ergs\,cm}^{-3}$ (Habing field), scaled by a factor $\chi$.  
For a detailed description of the treatment of the UV radiation field, see Appendix C of \citet{lepetit2006}.

We apply the model to pointing \pdr\ since it is the one that better approximates the model assumptions regarding geometry and stellar content, and it was set to assume a constant density in the slab.
We produced models varying the parameters representing the slab depth ($\av$), the atomic Hydrogen initial density ($\nh$), and the UV field radiation strength ($\chi$), while leaving the remaining parameters in their default settings, as detailed in \citet{lepetit2006}.

We run a grid of models varying $\log \chi$ in the range $[1,6]$ in steps of $1$, $\log \nh$ from $3$ to $6$ in steps of $1$, and $\av$ from $\sim0.5$\,mag to $\sim500$\,mag, in tenfold increases.
We used the line ratios \mbox{\cicotr\,=\,$\sum$[CI]/$\sum$CO} and \mbox{\cir\,=\,[CI]($ 2-1 $)/[CI]($ 1-0 $)} to determine the best solution for $\nh$ and $\chi$. \cicotr\ is indicative of the relative importance of the two main species ([CI] and CO) with regard to the cooling mechanisms. \cir\ is indicative of the temperature at the depth in the slab where neutral carbon is located and is emitting.

\begin{figure}
	\includegraphics[width=0.5\textwidth]{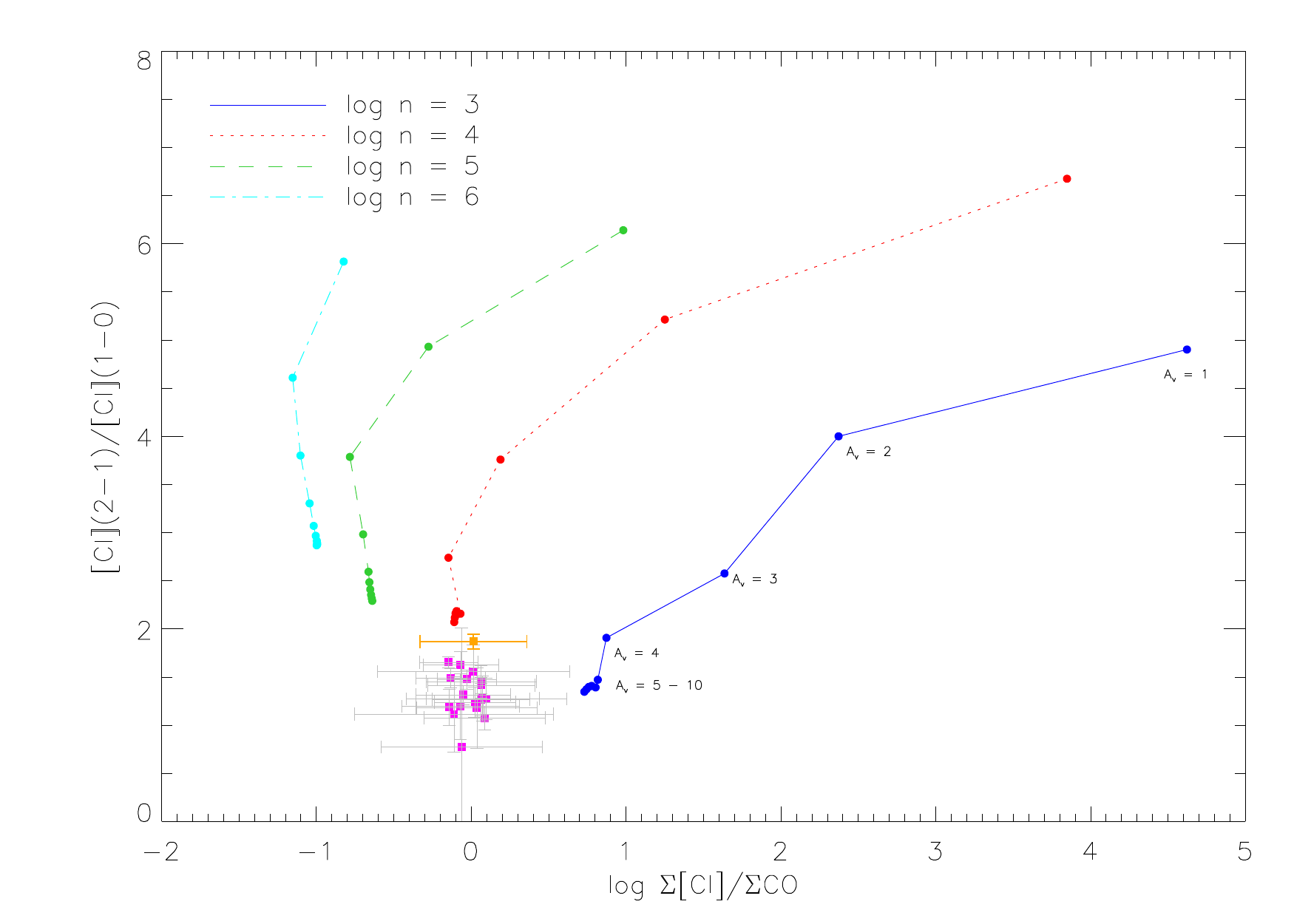}
	\caption{Behavior of $\nh$ for different values of $\av$, from 1 to 10 (filled big square) by steps of 1, for models with $\chi=10^{4}$ plotted against the \cir\ and \cicotr\ ratios. The line ratios observed at each SLW pixel of pointing \pdr\ are plotted with their error bars. We see that the simulated values closest to the observed ratio are the ones for models with $\nh = 10^{3-4}$ and $\av > 4-5$.}
	\label{fig-modres}
\end{figure}

First, we compare the ratio against the corresponding line ratio estimated by the model as a function of $\nh$ and $\chi$, obtaining the most likely solution $\nh \sim 10^3$; $\chi \sim 10^4$.
We then test these results, running a grid of models this time with $\chi=10^{3}$ fixed, and varying $\nh$ and $\av$. The range of densities is the same as before, while $\av$ ranges between 1 and 10 in unity steps.
Taking into account the relationship between $\Nh$ and $\av$ (e.g., \citealt{bohlin1978,rachford2002,lepetit2006}),

\begin{eqnarray}
\av & \sim & 5.34\times10^{-22} \left[\frac{\Nh}{{\rm cm}^{-2}}\right] \label{eq-avcd} \\
& \sim & 1.65\times 10^{-3}\left[\frac{n_{H}}{{\rm cm}^{-3}}\right]\left[\frac{\Delta pdr}{{\rm pc}}\right], \label{eq-av}
\end{eqnarray}

\noindent this means that, by Eq. \ref{eq-av}, the width of the slab varies between $\dpdr \sim 6\times 10^{-4}$\,pc (for $\log \nh=6$; $\av=1$) and $\dpdr \sim 6$\,pc (for $\log \nh =3$; $\av =10$).
This is illustrated in Fig. \ref{fig-modres}. Each of the lines represent the locus of points for a given density and varying slab width in the [\cir ;\cicotr] space. The square is the ratio obtained from the data of the central pixel of pointing \pdr, plotted with its respective error bars. We see that the simulated values closest to the observed ratio are the ones for models with $\nh = 10^{3-4}$ and $\av = 4-5$.

From the model then, we obtain the values $\chi\sim10^{4}$, $\nh\sim10^{3}$ and $\dpdr$ between $0.25$ and $3.0\,$pc (with Eq. \ref{eq-av}).

\subsubsection*{Estimation of $\chi$ at the PDR surface}

In order to estimate the UV radiation field impacting on the PDR and to compare it with the value derived from the model we integrated the spectrum of the O8V ionizing star of \mia\ \citep{martins2010} between 912\,\AA\ and 2400\,\AA, following Appendix C of \citet{lepetit2006}. The star spectrum and radius ($R_{*}\sim 8.166\,$R$_{\odot}$) were computed and facilitated by F. Martins (priv. comm.). 
The distance of the PDR surface from the ionizing star was measured on the SPIRE 350$\mm$ image and was estimated at $d\sim 2.15\,$pc. A simple dilution factor $(R_{*}/d)^{2}$ was applied and the backscattering of the radiation by dust at the PDR surface was taken into account, as recommended by \citet{lepetit2006}. We obtain a value of $\sim 1925$ in Habing units. This value is in good agreement with the $\sim 1000$ derived from our measurements using the PDR code, especially if we consider that dust present in the ionized region, as shown by the 24$\mm$ emission \citep{martins2010}, can absorb part of the radiation and diminishes that reaching the PDR surface.

\section{Discussion}
\label{sec-disc}

Combining the results from sections~\ref{sec:radex} and \ref{sec-model}, we see that the physical parameters corresponding to \mia\ are most likely between those derived with RADEX with $\nh=10^{4}-10^{6}\,$cm$^{-3}$, while for the diffuse regions \pdr\ and HII the density might be lower.
Assuming a standard \co\ abundance \mbox{${\rm ^{12}CO/H_{2}}\sim8\times10^{-5}$} \citep{frerking1982,roellig2011} and considering that \mbox{$N_{{\rm H}} = N({\rm H})+2N({\rm H_{2}})$}, we obtain the total hydrogen column density values show in columns 3 and 4 of Table~\ref{table-Nh}.

On the other hand, using the column density map obtained by \citet{anderson2012} for \mia, we averaged their column density values obtained at the position of the different pointings within the beam of the central pixel of the SLW array ($\sim 36\arcsec$ at $350\mm$, \citealt{makiwa2013}), obtaining the total hydrogen column densities shown in column 2 of Table~\ref{table-Nh}.

For pointings C1, C2, and IRS1 it was possible to obtain a more accurate value for $ \Nh $ with RADEX and not just a lower limit. 
Among these three positions, only for C1 is the density value obtained with \citet{anderson2012} map between the RADEX values. Nevertheless, the column densities derived by both methods are, in a first order, in agreement. The assumptions on the dust properties play a crucial role when deriving the column density from dust observations, therefore we cannot make a detailed comparison.

\begin{table}
	\renewcommand{\arraystretch}{1.2}
	\centering
	\caption{$ \Nh $ obtained for the different pointings.}
	\label{table-Nh}
	\begin{tabular}{l|c|ccc}
		\hline \hline
		 & \multicolumn{4}{c}{$ \Nh\,(\times10^{20}\,{\rm cm}^{-2}) $} \\
		 \cline{2-5}
		Pointing & Anderson+\citeyear{anderson2012} & \multicolumn{3}{c}{RADEX ($\mathrm{n_{H_{2}}}$ in cm$ ^{-3} $)} \\
		 & $ \Nh $ map & $ 10^{6}$ & $ 10^{5}$ & $10^{4}$ \\
		\hline
		 C1 & $ 440 $ & $ 250 $ & $ 250 $ & $ 750 $ \\
		 C2 & $ 110 $ & $ 250 $ & $ 250 $ & $ 500 $ \\
		 C6 & $ 100 $ & $ >2.5 $ & $ >1.25 $ & $ >2.5 $ \\
		 C8 & $ 78 $ & $ >7.5 $ & $ >7.5 $ & $ >12.5 $ \\
		 \pdr & $ 12 $ & \ldots & \ldots & \ldots \\
		 HII & $ 13 $ & \ldots & \ldots & \ldots \\
		 IRS1 & $ 51 $ & $ 175 $ & $ 175 $ & $ 500 $ \\
		 IRS2 & $ 34 $ & $ >5 $ & $ >5 $ & $ >12.5 $ \\
		 OFF & $ 28 $ & $ >7.5 $ & $ >7.5 $ & $ >12.5 $ \\
		 \hline \hline
	\end{tabular}
\end{table}

For the other pointings, with RADEX we only obtain lower limits to the column density, and correspondingly, the average values from \citet{anderson2012} map are higher than the RADEX values. In particular, pointing C6 shows a large difference between the two estimations, in line with the suggestion made in section~\ref{sec:radex} that its temperature is lower than that obtained with RADEX.

Any of the two methods show the differences in gas properties throughout \mia. The similar values for average column density of pointings \pdr\ and HII suggest that the PDR of \mia\ also extends on the plane-of-the-sky direction, supporting the bubble-shaped morphology proposed for it.

Pointings C6 and C8 show temperatures and densities intermediate between the warm and dense regions on the PDR, and its diffuse and cooler parts. 
As mentioned in section~\ref{sec:radex}, the relatively high temperature given by the code for pointing C6 is likely overestimated, leading to a slightly larger densities.

\section{Summary}
\label{sec-sum}

We have obtained \herschel\ SPIRE-FTS spectra towards 9 positions in the \mia\ \hii\ region, detecting the [CI] lines at $370$ and $609\mm$, the $205\mm$ [NII] transition, the \co\ ladder between the $J=4$ and $J=13$ levels and the \tco\ ladder between the $J=5$ and $J=14$ levels. CH$ ^{+} $ was detected in absorption at all positions in the region, however the low spectral resolution of the spectra do not allow us to obtain quantifiable information from that line. 
Nevertheless, its ubiquitous absorption suggest the presence of diffuse gas along the line of sight, while \mia\ may emit in this line with varying absorption depth.
The [NII] emission line is strong and is also detected over the entire field. The [CI] lines are detected in almost all detectors with a ratio which shows little variation throughout the region. This suggests the presence of low-density PDR over the entire RCW 120 region.
This is further supported by the temperatures obtained with the ratio of the two [CI] lines detected, which show little variation throughout the region.

The low-excitation $^{12}$CO lines are detected everywhere, while higher-excited lines are only detected in the condensations C1, C2 and IRS1 together with $^{13}$CO. We use RADEX to derive the physical properties at these positions. The gas temperatures are $45-250\,$K for densities of $10^4-10^6\,{\rm cm}^{-3}$, and a high column density that is in agreement with dust analysis. 
The excited CO could arise either from the edge of the dense irradiated structure or small dense clumps containing young stellar objects. We see the excited CO emission in several detectors, partly in an elongated region, coming from the PDR and/or several young stellar objects. 
The analysis of the other condensations C6 and C8 reveal a less dense medium with still high gas temperatures.
For the positions HII, CLay and OFF, where no condensations are observed, reveal the lowest densities with a highest \cicotr\ ratio.

We model the PDR of \mia\ (poiting CLay) with the Meudon PDR code. We obtain a hydrogen density of $\nh\sim 10^{4.3}$\,cm$^{-3}$ and an ionizing radiation field $\chi\sim10^{3}$ in Habing units.
The value for $\chi$ agrees with what is expected from the emission of an O8V star, as is the ionizing star of \mia.


\begin{acknowledgements}
We thank Dominique Benielli for assistance in the data calibration, Edward Polehampton for the useful discussion on the data calibration modes, and the referee for the helpful remarks that lead to a vastly improved version of this manuscript. We thank the French Space Agency (CNES) for financial support. J.A.R. acknowledges support by CNES on his post-doctoral fellowship.
\end{acknowledgements}

\def\aj{AJ}%
\def\araa{ARA\&A}%
\def\apj{ApJ}%
\def\apjl{ApJ}%
\def\apjs{ApJS}%
\def\ao{Appl.~Opt.}%
\def\apss{Ap\&SS}%
\def\aap{A\&A}%
\def\aapr{A\&A~Rev.}%
\def\aaps{A\&AS}%
\def\azh{AZh}%
\def\baas{BAAS}%
\def\jrasc{JRASC}%
\def\memras{MmRAS}%
\def\mnras{MNRAS}%
\def\pra{Phys.~Rev.~A}%
\def\prb{Phys.~Rev.~B}%
\def\prc{Phys.~Rev.~C}%
\def\prd{Phys.~Rev.~D}%
\def\pre{Phys.~Rev.~E}%
\def\prl{Phys.~Rev.~Lett.}%
\def\pasp{PASP}%
\def\pasj{PASJ}%
\def\qjras{QJRAS}%
\def\skytel{S\&T}%
\def\solphys{Sol.~Phys.}%
\def\sovast{Soviet~Ast.}%
\def\ssr{Space~Sci.~Rev.}%
\def\zap{ZAp}%
\def\nat{Nature}%
\def\iaucirc{IAU~Circ.}%
\def\aplett{Astrophys.~Lett.}%
\def\apspr{Astrophys.~Space~Phys.~Res.}%
\def\bain{Bull.~Astron.~Inst.~Netherlands}%
\def\fcp{Fund.~Cosmic~Phys.}%
\def\gca{Geochim.~Cosmochim.~Acta}%
\def\grl{Geophys.~Res.~Lett.}%
\def\jcp{J.~Chem.~Phys.}%
\def\jgr{J.~Geophys.~Res.}%
\def\jqsrt{J.~Quant.~Spec.~Radiat.~Transf.}%
\def\memsai{Mem.~Soc.~Astron.~Italiana}%
\def\nphysa{Nucl.~Phys.~A}%
\def\physrep{Phys.~Rep.}%
\def\physscr{Phys.~Scr}%
\def\planss{Planet.~Space~Sci.}%
\def\procspie{Proc.~SPIE}%
\let\astap=\aap
\let\apjlett=\apjl
\let\apjsupp=\apjs
\let\applopt=\ao

\bibliographystyle{aa}

\end{document}